\documentclass[12pt]{article}
\usepackage{cite}
\usepackage{amsbsy}
\usepackage{hyperref}
              
\title{A Theory of Gravitation Covariant under $Sp(4, \mathbf{R})$}
\author{M. Toller 
\thanks{Professor Emeritus of Trento University} 
\thanks{e-mail: toller@iol.it; URL: http://www.marco-toller.it/}\\ 
via Malfatti n. 8  \\
I-38100 Trento, Italy}
\begin{document}
\maketitle


\begin{abstract}
We present a Lagrangian theory of gravitation that develops some ideas proposed several years ago. It is formulated on the 10-dimensional space $\mathcal{S}$ of the local Lorentz frames (tetrads) and it is covariant under the symplectic group $Sp(4, \mathbf{R})$, locally isomorphyic to the anti-de Sitter group $SO(2, 3)$. The corresponding transformation formulas contain a constant $ \ell$, besides the light velocity $c$.  We also assume the covariance under the ``total dilatations'' of all the coordinates of the tangent spaces of $\mathcal{S}$. These symmetries, that we may call ``augmented Lorentz covariance'', are spontaneously broken and the corresponding (generalized) Goldstone fields, that we call ``augmentons'', behave as the components of a 5-vector of $SO(2, 3)$. Its square can be interpreted as the Brans-Dicke scalar field, that describes a variable gravitational coupling. The source of the augmentonic fields is provided by the Dirac fields. Finally, we discuss the physical relevance of the theory and its possible  further developments.


11.30.Qc (Spontaneous symmetry breaking);

\end{abstract}

\newpage
                
\section{Introduction}
\label{Introduction} 
Relativistic quantum fields defined on the Poincar\'e group (instead of the Minkowski spacetime) have been proposed by Lur\c cat in 1964 \cite{Lurcat}. The motivations came from strong interaction physics, where the presence of Regge trajectories suggested a ``dynamical role of spin''. 

It was argued that the relativistic angular momentum and the 4-momentum had to be treated on an equal footing, as well as the transformations generated by them, namely the Lorentz transformations and the spacetime translations. Note the analogy with the joint treatment of space and time (as well as momentum and energy) in relativistic theories.

In other words, all the elements of the Poincar\' e Lie algebra have to be treated in a symmetric (or impartial) way. We think that it is useful to give a name to this idea and we propose to call it the ``equity principle''.

It implies that, besides an upper bound to the velocity, there are upper bounds to the acceleration and to the angular velocity. The existence of a maximal acceleration has been discussed by Caianiello and many other authors \cite{Caianiello,CDFMV,Brandt1,Scarpetta,Brandt2,Papini,Toller16}. 

In the following decades these ideas have been applied to gravitation. It is interesting to note that a similar evolution, from strong to gravitational interactions, took place, about in the same period, in the field of the string theories \cite{CCCV}.

In the treatment of gravitational theories, based on the ideas of General Relativity (GR), the Poincar\' e group manifold has to be replaced by the fiber bundle \cite{KN,CB} of the local Lorentz frames (tetrads), that we indicate by $\mathcal{S}$. 

It is a principal fiber bundle in which the pseudo-Riemannian spacetime $\mathcal{M}$ is the base manifold and the orthochronous Lorentz group $O(1, 3)^{\uparrow}$ is the structure group, that acts freely and transitively on every fiber.

A fiber is composed of all the local Lorentz frames that have the same origin in $\mathcal{M}$ and it gives a mathematical description of the concept of ``spacetime coincidence'', indicated by Einstein \cite{Einstein2} as one of the fundamental concepts of GR.

We use the letters $i, j,\ldots, p, q,$ to represent indices that take the values $0, 1, 2, 3$ and for the Minkowskian metric tensor we adopt the convention $g_{00} = -1,  \, g_{11} = g_{22} = g_{33} =  1$.

The infinitesimal parallel displacements of the frames along the axes of a tetrad are described by four vector fields $A_i$ defined on $\mathcal{S}$ and the infinitesimal Lorentz transformations are generated by six vector fields $A_{[ik]} = - A_{[ki]}$ tangent to the fibers.

It is convenient to introduce also a more general notation in which $A_4 = A_{[32]}$, $A_5 = A_{[13]}$, $A_6 = A_{[21]}$ generate rotations around the spatial 4-vectors of the tetrad, $A_7 = A_{[01]}$, $A_8 = A_{[02]}$, $A_9 = A_{[03]}$ generate Lorentz boosts along the same spatial 4-vectors. We call the ten vector fields $A_{\alpha}$, with $\alpha = 0, 1,\ldots, 9$, the ``fundamental vector fields''.  We use a similar notation also for the indices of other quantities.

We also call them the ``fundamental derivations'' of the algebra of the smooth functions defined on $\mathcal{S}$.  They generate a 10-dimensional subspace $\mathcal{T}$ of the much larger vector space of all the derivations (or vector fields). The fields $A_i$ generate the ``horizontal'' subspace $\mathcal{T}_H \subset \mathcal{T}$, while the fields $A_{[ik]}$ generate the ``vertical'' subspace $\mathcal{T}_V \subset \mathcal{T}$.  Note that they have a direct  operational interpretation, at least in a macroscopic context \cite{Toller2,Toller1}.

We assume that the ten fundamental vector fields are linearly independent at all the points  $s \in \mathcal{S}$, namely they define a basis in every tangent space $\mathcal{T}_s$.  In this way we define an isomorphism between  every tangent space and $\mathcal{T}$. This means that  $\mathcal{S}$ acquires a structure of absolute  parallelism or teleparallelism. 

The commutators (or Lie brackets) of the fundamental derivations can be written in the form
\begin{equation} \label{Coeff}
[A_{\alpha}, A_{\beta}] = F_{\alpha \beta}^{\gamma}  A_{\gamma}.
\end{equation}
The functions $F_{\alpha \beta}^{\gamma} = - F_{\beta \alpha}^{\gamma}$, that can depend on $s$, are called the ``structure coefficients''. 
They satisfy the ``generalized Bianchi identities''
\begin{equation} \label{Jacobi}
A_{\alpha} F_{\beta \gamma}^{\delta} 
+ A_{\beta} F_{\gamma \alpha}^{\delta} 
+ A_{\gamma} F_{\alpha \beta}^{\delta} 
+ F_{\alpha \beta}^{\eta} F_{\gamma \eta}^{\delta} 
+ F_{\beta \gamma}^{\eta} F_{\alpha \eta}^{\delta} 
+ F_{\gamma \alpha}^{\eta} F_{\beta \eta}^{\delta} =  0,
\end{equation}
that follow from the Jacobi identity for the commutators.

If $\mathcal{S}$ is the bundle of Lorentz frames of a pseudo-Riemannian  spacetime $\mathcal{M}$, the horizontal subspaces of $\mathcal{T}_s$ define a connection, the vector fields $A_i$ represent the covariant derivatives, the quantities $- F_{ik}^{[jl]}$ and $F_{ik}^{j}$ are the components of the curvature and torsion tensors and, for some values of the indices, eq.\ (\ref{Jacobi}) gives the usual Bianchi identities. 

The structure of absolute parallelism of $\mathcal{S}$  can be used to describe more general, ``nonlocal''  geometries, in which $\mathcal{S}$ is not a fiber bundle. Then, the spacetime coincidence and the spacetime manifold $\mathcal{M}$ are not  defined any more and they may only be considered as approximate concepts.

These geometries may describe physical phenomena at very short distances, near to a fundamental length $\tilde \ell$, presumably of the order of the Planck length $\ell_P$.

We also introduce in the space $\mathcal{S}$ the ``fundamental 1-forms'' $\omega^{\beta}$, defined by
\begin{equation} \label{OneForms}
i_{\alpha} \omega^{\beta} = i(A_{\alpha}) \omega^{\beta} = \delta_{\alpha}^{\beta}.
\end{equation}
If $A$ is a vector field, we indicate by $i(A)$ the interior product operator acting on the differential forms. The exterior derivatives of these 1-forms are given by \cite{KN,CB}
\begin{equation} \label{Ext}
d \omega^{\gamma} = - 2^{-1} F_{\alpha \beta}^{\gamma} \, \omega^{\alpha} \wedge \omega^{\beta}.
\end{equation}
If $\mathcal{S}$  is a Lie group, the structure coefficients are constant,  the quantities $\omega^{\beta}$ are the Maurer-Cartan forms and eq.\ (\ref{Ext}) is the Maurer-Cartan equation.

The differential forms
\begin{equation} \label{Eta}
\eta = \omega^0 \wedge \omega^1 \wedge \omega^2 \wedge \omega^3 
= (24)^{-1} \epsilon_{ijkl} \, \omega^i \wedge \omega^j \wedge \omega^k \wedge \omega^l,
\end{equation}
\begin{equation} \label{EtaI}
\eta_i  = 6^{-1} \epsilon_{ijkl} \, \omega^j \wedge \omega^k \wedge \omega^l = i(A_i) \eta, \qquad
\omega^k \wedge \eta_i = \delta^k_i \eta
\end{equation}
appear in many formulas.

The fundamental forms $\omega^{\alpha}$ can be used as the dynamical variables of a theory of gravitation. The structure coefficients, or the exterior derivatives $d \omega^{\alpha}$, play the role of the derivatives of these dynamical variables.

Around the year 1978, one finds a renewed attention for physical theories based on this kind of geometry \cite{Smrz,NR1,NR2,Toller3,TollerVanzo,CSTVZ}. Besides the gravitational field, also the Klein-Gordon and the Dirac fields have been considered. A generalization, based on a larger structure group, has also been used to describe ``internal'' gauge fields, as the Maxwell and the Yang-Mills fields \cite{CSVZ}.  Other treatments of these and similar problems have appeared more recently, see for instance refs.\ \cite{Helein, HeleinVey}.

An important improvement has been proposed \cite {NR1,Toller3}, namely the action principle was defined by an integral 
\begin{equation} \label{ActionPrinciple}
\delta \int_S \lambda = 0,
\end{equation}
of a differential 4-form $\lambda$, called the ``Lagrangian form'', over an arbitrary 4-dimensional compact submanifold $S$ of $\mathcal{S}$ with a boundary $\partial S$, on which the condition $\delta \omega^{\alpha} = 0$ is required. A conserved quantity is defined by a closed 3-form, to be integrated over a 3-dimensional submanifold. 

The Lagrangian form may depend, in accord with the rules of coordinate-free differential geometry, on $\omega^{\alpha}$, on $F_{\alpha \beta}^{\gamma}$, on the matter fields $\Psi_U$ and on their derivatives $A_{\alpha} \Psi_U$. In the following we consider simpler Lagrangian forms in which the quantities $F_{\alpha \beta}^{\gamma}$ and $A_{\alpha} \Psi_U$ are only contained implicitly in the exterior derivatives (\ref{Ext}) and
\begin{equation} \label{Ext1}
d \Psi_U = A_{\alpha} \Psi_U \, \omega^{\alpha}.
\end{equation}
The corresponding field equations do not contain derivatives of the structure coefficients and second order derivatives of the matter fields and we say that we are dealing with a ``first order formalism''.

Ne'eman and Regge (NR) in refs.\ \cite{NR1,NR2} have described the Einstein-Cartan (EC) theory of gravitation \cite{HHKN} by means of a Lagrangian form that, with our notations, is given by
\begin{equation} \label{LagGM}
\lambda = \lambda^G + \lambda^M,
\end{equation}
\begin{equation} \label{NR} 
\lambda^G = (32 \pi G)^{-1} \epsilon_{ikjl} \, (d \omega^{[ik]} + g_{mn} \, 
\omega^{[im]} \wedge \omega^{[nk]}) \wedge \omega^j \wedge \omega^l - (8 \pi G)^{-1} \Lambda \eta, \quad
\end{equation}
where $\lambda^G$  is the gravitational (or geometric) Lagrangian form, $\lambda^M$ describes matter and $G$ is Newton's gravitational constant. We have added a term with the cosmological constant $\Lambda$, that in the meantime has acquired a considerable importance.

It is important to remark that from this Lagrangian one obtains, besides the  EC equations that involve curvature and torsion,  other equations that fix the other structure coefficients in the way required by the structure of a bundle of Lorentz frames.

A different, more complicated, Lagrangian form was proposed in ref.\ \cite{Toller3}, but it was abandoned when the simpler Lagrangian (\ref{NR}) appeared.

The EC theory, that is not, at present, experimentally distinguishable from GR, takes into account both the energy-momentum and the spin of matter and  both the curvature and the torsion of the spacetime manifold $\mathcal{M}$. It can be considered as a first step in the direction suggested by the equity principle.  It has initiated the development of several gauge theories of gravitation, as it is explained in ref.\ \cite{BH}.

A detailed general discussion of the field equations and of the conservation laws (Noether theorem) of the theories formulated in the space $\mathcal{S}$  is given in ref.\ \cite{CSTVZ} and some results are summarized in the next Section \ref{General}.

In the same article it has been shown that also the Brans-Dicke (BD) scalar field \cite{BD,Brans}, that describes a variable gravitational coupling, can be defined in terms of the geometric properties of the space $\mathcal{S}$. Variations of the gravitational constant have not yet been observed \cite{MU}, but the introduction  of  this scalar field was motivated  by Dirac's large number problem \cite{Dirac} and by the ideas of Mach \cite{Mach}  about the influence of very far celestial bodies on the locally observed phenomena. Another motivation, based on covariance properties, is discussed in Section \ref{Revisited}.

The theory described in ref.\ \cite{CSTVZ}, that we call  the ``geometrized BD  (GBD) theory'',  also includes some aspects of the EC theory, namely it takes into account torsion and spin.  Scalar-tensor theories with torsion have been considered by several authors, see for instance ref.\ \cite{Aldersley}.

More specifically, in the GBD formalism, the BD  field 
\begin{equation} \label{GBD}
\Phi^{BD} = (8 \pi G)^{-1} = \phi^2, 
\end{equation}
that represents the inverse of the locally measured rationalized (variable) gravitational coupling, is given as a function of the structure coefficients by the formula
\begin{equation} \label{PhiDef}
\phi = (12)^{-1} g^{kj} F_{[ik]j}^i.
\end{equation}

Both the EC and the GBD theories assign  a physical meaning to a larger part of the 450 structure coefficients, besides the 36  components of curvature, that are the only variable structure coefficients in GR. In the following, we try to proceed further in this direction.  There are still many unexploited degrees of freedom in the absolute parallelism structure of $\mathcal{S}$.

Actually, the GBD theory described in ref.\ \cite{CSTVZ} contains a free parameter $m$. We choose the value $m = 4$, because in this case the gravitational Lagrangian is invariant under ``total dilatations''  acting on all the ten coordinates of the space $\mathcal{T}$.  This symmetry is spontaneously broken, because  the field $\phi$, that we call the ``dilatonic field'', is  not invariant and it  takes a non vanishing constant value in a vacuum state. 

The theories we shall consider are symmetric under all the diffeomorphisms of the space $\mathcal{S}$ and also under a covariance group,  that acts linearly on the indices of the fields.  We consider only global covariance transformations, that do not depend on the point $s \in \mathcal{S}$. 

Local (gauge) covariance transformations depending on $s$ would spoil the structure of absolute parallelism of $\mathcal{S}$ and its operational interpretation.  They  can be treated as diffeomorphisms of  $\mathcal{S}$, possibly with increased dimension $n > 10$  \cite{CSVZ}.

In a local theory, the Lorentz group has a double interpretation, namely as the structure group of the principal bundle, that is  a sugroup of its diffeomorphism group, and as a subgroup of the covariance group.  In the nonlocal theories, the first interpretation is lost.

 In general, a theory has covariance transformations that affect the internal indices of the matter fields, for instance color $SU(3)$ or weak $SU(2)$.  However, now we are interested in geometric transformations that operate only on the spin indices and on the index $\alpha$ that labels the fundamental fields  $A_\alpha$.
 
The geometric covariance group $\mathcal{C}$ of GR and of the EC theory is the orthochronous Lorentz group $O(1, 3)^{\uparrow}$. It acts on the space $\mathcal{T}$, namely on the vector fields $A_{\alpha}$, by means of the direct sum of its vector and  antisymmetric tensor representations, that operate, respectively, on the horizontal and the vertical subspaces. 

A natural way to implement the equity principle is to introduce a larger ``augmented'' covariance group $\mathcal{C}$ that acts on the space $\mathcal{T}$ by means of an irreducible representation. Then, the horizontal and the vertical subspaces cannot any more be defined in an invariant way.

The comparison of horizontal and vertical vectors requires the introduction of a new fundamental constant  $\ell$, that in theories with a constant gravitational coupling is a length $\tilde \ell$.   

A different alternative way to introduce a fundamental length has been discussed by many authors \cite{Amelino1,Amelino2,MS,Hossenfelder1,Hossenfelder2,Toller21,Toller22} and is based on a ``deformation'' of the Lorentz group. We follow a different approach, namely we ``augment'' the Lorentz covariance, rather than deforming it. The augmented covariance has to be spontaneously broken, but some of its consequences, as the conservation laws, survive.

A general discussion of the possible augmented covariance groups was given in ref.\ \cite{Toller4}. In this analysis the most relevant group, locally isomorphic to $GL(4, \mathbf{R})$, was inexplicably omitted and an erratum was published to correct this mistake.

The same article proposed two Lagrangian forms invariant under the anti-de Sitter group $SO(2, 3)^c$ (the superscript ``c'' indicates the connected component of the unit). One of them was also invariant under total dilatations.

Unfortunately, the field equations obtained from these Lagrangians had no interesting solution, besides the solutions of the EC theory and the solutions obtained from them by means of covariance transformations. The aim of the present article is to present (after a long time) a modified Lagrangian with more interesting solutions.

In the next Section \ref{General} we briefly recall the general form of the field equations and of the conservation laws. In Section \ref{Revisited} we describe a new approach to the GBD theory, that provides an useful guide for the construction of more general models with augmented covariance. 

In Section \ref{Groups} we describe and justify the ``fully augmented'' covariance group $GL(4, \mathbf{R})$ and in  Section  \ref{Spinor} we consider some properties of this group and of its simplest finite-dimensional representations.  In Section \ref{Partially} we discuss the difficulties encountered in the construction of a theory with fully augmented covariance and we introduce  a ``partially augmented'' covariance group $\mathcal{C} \subset GL(4, \mathbf{R})$ containing a symplectic subgroup  $Sp(4, \mathbf{R})$, the space reflection $\gamma_0$ and the total dilatations. In Section  \ref{Vacuum} we describe  the vacuum states of a theory with this covariance. 

In Section \ref{Model} we propose a specific Lagrangian with partially augmented covariance and we derive the field equations.  In Section \ref{Conservation}  we complete the treatment discussing the conservation laws following from the augmented covariance.  In Section \ref{Foliation} we  present some geometric properties of the space $\mathcal{S}$ that follow from the field equations.   Finally, in Section \ref{Conclusions} we discuss  some possible future developments.

\section{The Lagrangian formalism in the space $\mathcal{S}$}
\label{General} 
   
In the present Section we summarize the general formalism that we use to derive the field equations and the conservation laws from the action principle (\ref{ActionPrinciple}).  We follow, with some simplifications, the treatment of ref. \cite{CSTVZ}.
 
If the quanties $F_{\alpha \beta}^{\gamma}$ and $A_{\alpha} \Psi_U$ appear only implicitly through the exterior derivatives $d \omega^{\gamma}$ and $d \Psi_U$, one can define the 2-forms $\sigma_{\gamma}$ and the 3-forms $\pi^U$ with the properties
\begin{equation} \label{Conjugated}
\frac{\partial \lambda}{\partial F_{\alpha \beta}^{\gamma}} 
= - 2^{-1} \omega^{\alpha} \wedge \omega^{\beta} \wedge \sigma_{\gamma}, \qquad
\frac{\partial \lambda}{\partial A_{\alpha} \Psi_U} = \omega^{\alpha} \wedge \pi^U.
\end{equation}
A comparison with the usual Lagrangian formalism suggests to call the forms $\sigma_{\gamma}$ and $\pi^U$ the ``canonically conjugate forms'' corresponding to the fields $\omega^{\gamma}$ and $\Psi_U$. 

The same equations can be obtained from Lagrangians that depend on $F_{\alpha \beta}^{\gamma}$ and $A_{\alpha} \Psi_U$ in a more general way, but then we have to impose some ``normal'' field equations \cite{CSTVZ}, that in the simple case we are considering are automatically satisfied.

From a variation of the fields $\omega^{\alpha}$, and $\Psi_U$, the action principle gives the ``tangential'' equations
\begin{equation} \label{Tangential1}
d \pi^U = \frac{\partial \lambda}{\partial \Psi_U},
\end{equation}
\begin{equation} \label{Tangential2}
d \sigma_{\alpha} = - \tau_{\alpha},
\end{equation}
where
\begin{equation} \label{TauAlpha}
\tau_{\alpha} = \tau(A_{\alpha}) = F_{\alpha \beta}^{\gamma} \, \omega^{\beta} \wedge \sigma_{\gamma}  - A_{\alpha} \Psi_U \pi^U + i_{\alpha} \lambda.
\end{equation}

It follows that 
\begin{equation} \label{Conservation1}
d \tau_{\alpha} = 0,
\end{equation}
namely the 3-forms $\tau_{\alpha}$ describe the density and the flow of quantities that are conserved as a consequence of the generalized Gauss law (\ref{Tangential2}). Their integrals
\begin{equation} \label{TenMomentum}
P_{\alpha} = \int_{\Sigma} \tau_{\alpha} = - \int_{\partial \Sigma} \sigma_{\alpha}
\end{equation}
on a suitable three-dimensional submanifold $\Sigma$ with boundary $\partial \Sigma$ are interpreted as the components of the 4-momentum and of the relativistic angular momentum, that we may call the components of the ``10-momentum''.  

The 10-momentum contains contributions from matter and from the gravitational fields. The last equality in eq.\ (\ref{TenMomentum}) shows that these quantities vanish if $\Sigma$ is a closed surface, for instance a spacelike slice in a spatially bounded universe. These conservation laws follow also from the Noether theorem and the symmetry of the theory under the diffeomorphisms of $\mathcal{S}$ generated by $A_{\alpha}$, independently of any covariance property.  Sometimes, it is useful to consider $\tau$ as a 3-form defined on $\mathcal{S}$ that takes its values in the dual $\mathcal{T}^*$ of the vector space $\mathcal{T}$.

If, as in the examples  we shall consider,  the Lagrangian depends linearly on the forms $d \omega^{\alpha}$, we can write
\begin{equation} \label{LambdaS}
\lambda = d \omega^{\alpha} \wedge \sigma_{\alpha} + \overline\lambda
\end{equation}
and the  equation (\ref{TauAlpha}) takes the simpler form
\begin{equation} \label{TauAlphaS}
\tau_{\alpha} = d \omega^{\eta} \wedge i_{\alpha} \sigma_{\eta} - A_{\alpha} \Psi_U \pi^U + i_{\alpha} \overline\lambda.
\end{equation}

A theory with covariance transformations also has other conservation laws. Consider an infinitesimal transformation parametrized by the infinitesimal parameter $\zeta$ and described by the operator X, which is a derivation of the algebra of fields, acting on the field indices, but not on their argument $s \in \mathcal{S}$, namely
\begin{equation} \label{InfCov}
\delta \omega^{\alpha} = \zeta X \omega^{\alpha} = \zeta X^{\alpha}_{\beta}  \omega^{\beta} , \qquad
\delta \Psi_U = \zeta X \Psi_U = \zeta X_U^V \Psi_V.
\end{equation}

If it does not affect the Lagrangian form, namely we have
\begin{equation} \label{Invariance}
\delta \lambda = \zeta X \lambda = 0,
\end{equation}
the corresponding conserved 3-form $\theta(X)$ is
\begin{equation} \label{Conserved}
\theta(X) = X \omega^{\alpha} \wedge \sigma_{\alpha} + X \Psi_U \pi^U.
\end{equation}
It may be considered as a  3-form defined on $\mathcal{S}$ that takes its values in the dual $L(C)^*$ of the Lie algebra of the group $C$ of the covariance transformations.

In fact, from eqs.\ (\ref{Conjugated}) we obtain, after some calculations, 
\begin{eqnarray} \label{Invariance2}
X \lambda = d \theta(X) + X \omega^{\alpha} \wedge (d \sigma_{\alpha} + \tau_{\alpha})
+ X \Psi_U \left( \frac{\partial \lambda}{\partial \Psi_U} - d \pi^U \right) = 0
\end{eqnarray}
and, if all the field equations are satisfied, we obtain a proof of the conservation law $d \theta(X) = 0$. 

The corresponding integrated conserved quantities are
\begin{equation} \label{ConservedQ}
Q(X) = \int_{\Sigma} \theta(X).
\end{equation}
Note that they do not necessarily vanish when $\Sigma $ is a closed surface and they depend only on the homology class of $\Sigma$.  If $\mathcal{S}$ has the topology of the Poincar\'e goup, the third homology group is generated by a single element corresponding to the rotation subgroup $SO(3)$. In a spatially bounded universe, for instance if $\mathcal{S}$ has the topology of the de  Sitter group,  a spacelike slice provides another generator of the homology group. 

If we choose an homology class, $Q$ is a function of the state of the system that takes its values in $L(C)^*$.  It is called the ``moment map'' \cite{Souriau,Woodhouse}. 

In some cases, it is possible, and convenient, to use eq.\ (\ref{Invariance2}) to replace some of the field equations (\ref{Tangential1}) with some conservation laws, which are simpler and have a more clear physical interpretation. We shall exploit this possibility in Sections \ref{Revisited} and \ref{Model}.

The contribution 
\begin{equation} \label{ThetaM}
\theta^M(X) =   X \Psi_U \pi^U
\end{equation}
of some matter fields  $\Psi_U$ (not necessarily all) is not, in general, conserved, but it satisfies a balance equation. We assume that the fields $\Psi_U$  are minimally coupled with the other fields, namely that  $\Psi_U$  and $d \Psi_U$  appear only in an invariant part $\lambda^M$ of the Lagrangian that does not contain derivatives of the other geometric and matter fields. Then, from the invariance of $\lambda^M$ and from the field equations we have
\begin{equation} \label{BalanceTheta}
d \theta^M(X) = - X\omega^{\alpha} \wedge \tau^M_{\alpha}
- X \hat \Psi_V \frac{\partial \lambda^M}{\partial \hat \Psi_V},
\end{equation}
where $\hat \Psi_V$ are the other matter fields that appear in $\lambda^M$ and
\begin{equation} \label{TauM}
\tau^M_{\alpha} =  - A_{\alpha} \Psi_U \pi^U + i_{\alpha} \lambda^M
\end{equation}
describes the 10-momentum of the matter fields $\Psi_U$.

\section{The GBD theory revisited}
\label{Revisited} 
In ref.\ \cite{CSTVZ} the GBD theory is described by a gravitational Lagrangian  depending only on the forms $\omega^{\alpha}$ and on the structure coefficients. The scalar field $\phi$ that appears in this Lagrangian is defined from the beginning  as a function of the structure coefficients given by eq.\ (\ref{PhiDef}).  Note that this is not a Lagrangian of the simple kind considered in Section \ref{General}.

It is clear that the vector fields  $A_{[ik]}$ used to define the variable structure coefficients that appear in eq.\ (\ref{PhiDef}) do not describe directly the usual infinitesimal Lorentz transformations, that can be represented by the scaled  fields
\begin{equation}  \label{Scaled}
\tilde A_{[ik]}  = \phi^{-1} A_{[ik]}, \qquad \tilde \omega^{[ik]}  = \phi \, \omega^{[ik]}.
\end{equation}
From the operational point of view, it is not clear which fields can more naturally be defined in terms of physical operations, that could involve gravitational forces.

There is no problem from a macroscopic point of view, but a rescaling of the fundamental vector fields by means of variable factors that represent measurable quantities is not justified from the operational point of view when a minimal time plays a relevant role.   In fact, the operational definition of a finite transformation generated by a scaled vector field requires multiple successive measurements of the variable fields that appear in the scaling formula and each measurement takes a non vanishing time. 
 
The GBD Lagrangian of ref.\ \cite{CSTVZ} is invariant under the infinitesimal total dilatations generated by the derivation $X_{td} $ defined by
\begin{equation} \label{TotalDil}
X_{td} A_{\alpha} =  A_{\alpha},  \qquad 
X_{td} \, \omega^{\alpha} = - \omega^{\alpha},   \qquad X_{td} \, \phi = \phi.
\end{equation}
For the scaled fields we have the transformation formulas
\begin{equation} \label{TotalDil2}
X_{td}  A_i =   A_i, \qquad X_{td} \tilde A_{[ik]} = 0, \qquad 
X_{td} \, \omega^i = -  \omega^i, \qquad  X_{td} \, \tilde \omega^{[ik]} = 0,   
\end{equation}
that describe spacetime dilatations (acting also on $\phi$), that involve only the horizontal subspace $\mathcal{T}_H \subset \mathcal{T}$  in disagreement with  the equity principle.  This is not a problem since, in any case, this principle is not satisfied by the GBD theory and a treatment based on the scaled variables (\ref{Scaled}) is mathematically simpler.

In the spacetime formalism, in the absence of the scalar field $\phi$, a gravitational Lagrangian invariant under  (global) spacetime dilatations must depend on the square $R^2$ of the curvature. Theories of this kind have been considered by various authors (see for instance  ref. \cite{Buchdahl,Starobinsky,RCVZ}) and have some problems with the Newtonian limit.  These problems can be avoided by introducing a scalar field, as explained in ref.\ \cite{RV}, where other references can be found.  One gives in this way an additional motivation for the BD field,  besides the ones cited in the Introduction.

In the present article, we use the GBD theory as a first step in the construction of more general theories with augmented covariance, in which the equity principle is satisfied.  For this reason, at the end of the present Section we have to reintroduce  the original variables $A_{[ik]}$, that transform according to eq.\ (\ref{TotalDil}) under total dilatations.

In the following, we present an alternative treatment of the  GBD theory proposed in ref.\ \cite{CSTVZ}, that is more suitable for the generalizations that we have in mind. We use provisionally the scaled variables, we consider the field $\phi$ as an independent dynamical variable and we treat it in the same way as the matter fields  (but separately from them). As we shall see, its geometric nature appears when we use the original fields $A_{\alpha}$.

We work in a first order formalism and, in order to obtain a second order equation for the field $\phi$, we have to introduce another independent 4-vector field $\beta_i$. A theory of this kind can be described by the geometric Lagrangian
\begin{equation} \label{GBDLag}
\lambda^G =  2^{-2} \phi^2 \epsilon_{ikjl} (d \tilde \omega^{[ik]} + g_{mn} \, \tilde \omega^{[im]} \wedge \tilde \omega^{[nk]}) \wedge \omega^j \wedge  \omega^l + \lambda^+ ,
\end{equation}
where 
\begin{equation} \label{GBDLagPlus}
\lambda^+ =  - \kappa \phi \beta^i d \phi \wedge \eta_i 
+ 2^{-1} \kappa \phi^2  \beta_i \beta^i  \eta  - \hat \Lambda \phi^4 \eta .
\end{equation} 
Note that the forms (\ref{Eta}) and  (\ref{EtaI}) are not affected by the rescaling.

If we disregard $\lambda^+$, that describes the dynamics of the scalar field $\phi$, this is just the NR Lagrangian (\ref{NR}) with the coupling constant $8 \pi G$ replaced by the field $\phi^{-2}$ according to eq.\ (\ref{GBD}) and with $\omega^{[ik]}$ replaced by $\tilde \omega^{[ik]}$. The term containing $\hat \Lambda $, if $\phi$ is constant, becomes a cosmological term with $\Lambda = \phi^2 \hat \Lambda $.  If it has to account for the accelerated expansion of the universe, we must have $\hbar \hat \Lambda \approx 10^{-122}$ and we find a large number problem, that, following Dirac's argument \cite{Dirac}, could suggest the introduction of a new scalar field.  Alternatively, one has to find a different explanation of the properties of the red shift of the very distant supernovae \cite{Riess,Perlmutter}.

Since it is proportional to the square of the BD field, the term containing $\hat \Lambda$  looks like a mass term, but we shall see that it does not appear in the field equation (\ref{GBD5}) that determines the propagation of $\phi$.

We assume a minimal coupling of matter with geometry, namely that the matter Lagrangian  $\lambda^M$ does not contain structure coefficients, derivatives of $\phi$ and the fields $\beta_i$. The forms canonically conjugated to $\beta_i$ vanish and the corresponding field equations are
\begin{equation} 
\frac{\partial \lambda}{\partial \beta^i} = - \kappa \phi  d \phi \wedge \eta_i + \kappa  \phi^2 \beta_i  \eta = 0,
\end{equation}
 namely, if $\kappa \phi \neq 0$,
\begin{equation} \label{Beta}
\tilde A_{[ik]} \phi = 0, \qquad  A_i \phi = \beta_i \phi. 
\end{equation}

The other canonically conjugated forms are
\begin{equation}  
\sigma_i = 0, \qquad
\tilde \sigma_{[ik]} =  2^{-1} \phi^2 \epsilon_{ikjl} \, \omega^j \wedge \omega^l, 
\end{equation}
\begin{equation}  \label{PiPsi}
\pi^{\phi} =  - \kappa \phi \beta^i \eta_i.
\end{equation} 

The field equation (\ref{Tangential2}), together with eq. (\ref{TauAlphaS}) takes the form
\begin{equation} \label{GBD1}
- 2^{-1} \phi^2 \epsilon_{ijkl} (d \tilde \omega^{[jk]} +  g_{mn} \, \tilde \omega^{[jm]} \wedge \tilde \omega^{[nk]}) \wedge \omega^l =  T^{\phi j}_i \eta_j + \tau^M_i,
\end{equation}
\begin{equation} \label{GBD2}
- \phi^2 \epsilon_{ikjl} (d \omega^j + g_{mn} \, \tilde \omega^{[jm]} \wedge \omega^n) \wedge  \omega^l   =  \tilde T^{\phi j}_{[ik]}  \eta_j  + \tilde \tau^M_{[ik]},
\end{equation} 
where 
\begin{eqnarray} \label{EnergyPhi}
& T^{\phi j}_i = \kappa \phi^2 (\beta_i \beta^j - 2^{-1} \beta_l \beta^l \delta_i^j) - \hat \Lambda \phi^4 \delta_i^j, \nonumber \\
&\tilde T^{\phi j}_{[ik]} = -  2 \phi^2  (\beta_i  \delta_k^j  -  \beta_k  \delta_i^j ). 
\end{eqnarray}

The forms  $\tau^M_{\alpha} $, given by eq.\ (\ref{TauM}), describe the 10-momentum of matter (not including $\phi$) and we assume that they are given by the localized  expression
\begin{equation}  \label{Localized}
\tau^M_{\alpha} = T^j_{\alpha} \eta_j.
\end{equation}
Note that also the forms $\tau_{[ik]} = \tau(A_{[ik]})$ have to be rescaled, namely we have to put
\begin{equation}  \label{ScaledTau}
\tilde \tau_{[ik]}  = \phi^{-1} \tau_{[ik]}.
\end{equation}

By means of eq.\ (\ref{Ext}) we can write these field equations in the more explicit form
\begin{eqnarray}  \label{HorDiff2}
&\tilde F_{p[jl]}^{[ik]} = 0, \qquad 
\tilde F_{[jl][mn]}^{[ik]} = \hat F_{[jl][mn]}^{[ik]},  \nonumber \\
&\phi^2 (- \tilde F_{ik}^{[jk]} + 2^{-1} \delta^j_i \tilde F_{lk}^{[lk]})  = T^{\phi j}_i + T^j_i, 
\end{eqnarray}
\begin{eqnarray} \label{VerDiff2}
&\tilde F_{[jk]l}^i = \hat F_{[jk]l}^i, \qquad \tilde F_{[jk][mn]}^i = 0,  \nonumber \\
&\phi^2 (\tilde F_{ik}^j + \delta^j_i \tilde F_{kl}^l - \delta^j_k \tilde F_{il}^l)  = \tilde T^{\phi j}_{[ik]} + \tilde T^j_{[ik]}.
\end{eqnarray} 

We have indicated by  $\hat F_{\alpha \beta}^{\gamma}$  the structure constants of the Poincar\' e Lie algebra.  With our conventions, the non vanishing ones are given by
\begin{eqnarray} \label{StructConst1}
&\hat F_{[ik][jl]}^{[mn]} = \delta_i^m g_{kj} \delta_l^n - \delta_k^m g_{ij} \delta_l^n
- \delta_i^m g_{kl} \delta_j^n + \delta_k^m g_{il} \delta_j^n& \nonumber \\
&- \delta_i^n g_{kj} \delta_l^m + \delta_k^n g_{ij} \delta_l^m
+ \delta_i^n g_{kl} \delta_j^m - \delta_k^n g_{il} \delta_j^m,&
\end{eqnarray}
\begin{equation} \label{StructConst2}
\hat F_{[ik]j}^l = - \hat F_{j[ik]}^l = g_{kj} \delta_i^l - g_{ij} \delta_k^l.
\end{equation}  

The equations (\ref{HorDiff2}) and (\ref{VerDiff2}) are compatible with the structure of principal bundle and they can easily be interpreted as equations in the spacetime $\mathcal{M}$.  We obtain the equations of the the EC theory with the additional source terms $T^{\phi j}_i$ and $\tilde T^{\phi j}_{[ik]}$.  We see that, even in the absence of spin, if $\phi$ is not constant, a non vanishing torsion may appear. We also obtain the equation
\begin{equation} \label{Beta2}
\beta_i  = 3^{-1} \tilde F_{i k}^{k} + 6^{-1} \phi^{-2} \tilde T_{[ik]}^{k}.
\end{equation}  
If the spin of matter is due to Dirac fields, the last term in the right hand side vanishes. 

The eq.\ (\ref{Tangential1}) for the field $\phi$ is
\begin{eqnarray} \label{GBD3}
&d \pi^{\phi}  - \frac{\partial \lambda^G}{\partial \phi} = \frac{\partial \lambda^M}{\partial \phi}.
\end{eqnarray}
If we assume that the matter Lagrangian $\lambda^M$ too is invariant under total dilatations,  the form
\begin{equation} \label{ThetaDil}
\theta_{td} = \theta_{td}^G + \theta_{td}^M, \qquad
\theta_{td}^G =  \phi \pi_{\phi}  =  - \kappa \phi^2 \beta^i \eta_i
\end{equation}
is conserved and we see from eq.\ (\ref{Invariance2}) that the  conservation law for $\theta_{td}$ can be used to replace the field equation (\ref{GBD3}).  After some calculations, using eqs. (\ref{Coeff}),  (\ref{Beta}) and  (\ref{Beta2}), it can be written in the form
\begin{equation}  \label{GBD4}
d \theta_{td}^G = \kappa \phi^2 (-  A_i \beta^i + \beta_i \beta^i) \eta = - d \theta_{td}^M.
\end{equation}

The balance equation (\ref{BalanceTheta}) gives
\begin{equation} \label{DThetaTD}
d \theta_{td}^M =   T^i_i \eta  + 2^{-1} \tilde T^i_{[jk]}  \tilde \omega^{[jk]} \wedge \eta_i
- \phi \frac{\partial \lambda^M}{\partial \phi}
\end{equation}
and a comparison with eq.\ (\ref{GBD4}) gives
\begin{equation}  \label{GBD5}
\kappa \phi^2 (A_i \beta^i - \beta_i \beta^i)  =  T^i_i, 
\end{equation}
if we assume that
\begin{equation}  \label{CondMat}
\phi \frac{\partial \lambda^M}{\partial \phi} 
=  2^{-1} \tilde T^i_{[jk]}  \tilde \omega^{[jk]} \wedge \eta_i.
\end{equation}
The last equation is satisfied if, for instance, there is no spinning matter and $\lambda^M$  does not depend on $\phi$.  Note that $\hat \Lambda$ has disappeared from the field equation (\ref{GBD5}).

If $G$ and $\phi$ are constant,  the field equations (\ref{GBD1}),  (\ref{GBD2}) become the equations of the EC theory. The equation (\ref{GBD5}), however,  gives $T^i_i = 0$, a requirement not present in the EC theory. The constant $\hat \Lambda$  determines the vacuum states of the theory,  in which $T^i_{\alpha}$ vanishes and all the fields are constant.  The structure coefficients are the stucture constants of the Poincar\'e or the (anti)-de Sitter group  given by eqs.\ (\ref{StructConst1}), (\ref{StructConst2}) and 
\begin{equation} \label{StructConst3}
\hat F_{i k}^{[j l]}= - K (\delta_i^j \delta_k^l - \delta_k^j \delta_i^l), 
\end{equation}
with 
\begin{equation} 
K =  3^{-1} \hat \phi^2 \hat \Lambda.
\end{equation}

In order to  obtain the original BD equations \cite{BD}, we have to use eq.\ (\ref{GBD}) and to introduce the new fundamental fields
\begin{eqnarray} \label{NewFields}
&\breve A_{[ik]} = \tilde A_{[ik]}, \qquad 
\breve A_k = A_k - \beta^i \tilde A_{[ik]},   \nonumber \\
&\breve \omega^{[ik]} = \tilde \omega^{[ik]} + \beta^i \omega^k - \beta^k   \omega^i, \qquad  
\breve \omega^i =\omega^i,
\end{eqnarray}
that have the properties
\begin{equation} \label{NTanDiff3}
\breve A_{[ik]} \Phi^{BD} = 0, \quad \breve A_i \Phi^{BD} = 2 \beta_i \Phi^{BD}, \quad  
 \breve A_i \beta_k =  A_i \beta_k  + \beta_i \beta_k - \beta_j \beta^j g_{ik}.
\end{equation}
We disregard the spin of matter and we have
\begin{equation} 
\breve \tau^M_{[ik]} = \tilde \tau^M_{[ik]}  = \tau^M_{[ik]} = 0, \qquad 
\breve \tau^M_k =  \tau^M_k.
\end{equation}
The new structure constants $\breve F_{\alpha \beta}^{\gamma}$ can be computed by means of eq.\ (\ref{Coeff}) or  eq.\ (\ref{Ext}).

In terms of the new fields, eq.\ (\ref{GBD5}), after some calculations, also using eq.\ (\ref{Beta2}),  takes the form
\begin{equation}  \label{GBD6}
2^{-1} \kappa \breve A^i \breve A_i \Phi^{BD} =  T^i_i,
\end{equation}
that is just one of the BD equations if
\begin{equation} \label{KOmega}
\kappa = 4 \omega + 6,
\end{equation}
where $\omega$ is the constant introduced by BD.  This is a massless wave equation, that encourages the interpretation of the dilatonic field $\phi$ as the Goldstone field \cite{Weinberg6, Strocchi} corresponding to the spontaneous breaking of the covariance under total dilatations.

A comparison with astronomical measurements in the solar system gives a rather high lower bound on the dimensionless parameter $\omega$. Recent data from Cassini-Huygens spacecraft give $2^{-2} \kappa' \approx \omega > 4 \times 10^4$ \cite{Will}. 

In a similar way we modify the field equations \ (\ref{HorDiff2}) and  (\ref{VerDiff2}). By means of  the formulas 
\begin{eqnarray}  \label{FTilde}
&\phi \tilde F_{jl}^{[ik]} = \breve  F_{jl}^{[ik]}  + \breve  F_{jl}^i \beta^k - \breve  F_{jl}^k \beta^i 
+ (\breve  A_j - \beta_j)  \beta^i \delta_l^k - (\breve  A_l - \beta_l) \beta^i \delta_j^k \nonumber \\
&- (\breve  A_j - \beta_j)  \beta^k \delta_l^i + (\breve  A_l - \beta_l) \beta^k \delta_j^i
+\beta_p \beta^p (\delta_j^i \delta_l^k - \delta_j^k \delta_l^i), \nonumber \\
&\tilde F_{jl}^i = \breve  F_{jl}^i + \beta_j \delta_l^i  - \beta_l \delta_j^i,
\end{eqnarray}
 we obtain the equation
\begin{equation} \label{VerDiff3}
\Phi^{BD} (\breve F_{jk}^i + \delta^i_j \breve F_{kl}^l - \delta^i_k \breve F_{jl}^l) = \tilde  T^i_{[jk]} = 0,
\end{equation} 
that shows that, in the absence of matter spin, the redefined torsion $\breve F_{jk}^i$ vanishes. Under this assumption, we can write the last  field equation  in the simplified form
\begin{eqnarray}  \label{HorDiff3}
&\Phi^{BD} (- \breve F_{jk}^{[ik]} + 2^{-1} \delta^i_j \breve  F_{lk}^{[lk]}) 
- \breve A_j \breve A^i \Phi^{BD} + \delta^i_j \breve A_k \breve A^k \Phi^{BD} \nonumber \\
&- 2^{-2} (\kappa -6)  (\Phi^{BD})^{-1} (\breve A_j \Phi^{BD} \breve A^i \Phi^{BD} - 2^{-1} \breve A^k \Phi^{BD} \breve A_k \Phi^{BD} \delta_j^i)  \nonumber \\
&+ \hat \Lambda \phi^4 \delta^i_j = T^i_j.
\end{eqnarray}
This is the second BD equation and, in conclusion, we have seen that, in the absence of spinning matter, the GBD theory is equivalent to the original BD theory.

The terms in eq.\ (\ref{HorDiff3}) that contain the derivatives of $\Phi^{BD} $ can be shifted to the right hand side and considered as the contribution of the scalar field to the energy-momentum tensor. It has been remarked \cite{Faraoni1} that, as a consquence of the presence of second order derivatives, the energy is not any more positive.  

It may be convenient to skip the change of variables (\ref{NewFields}), to  formulate the theory by means of the equations  (\ref{HorDiff2}), (\ref{VerDiff2}) and  (\ref{GBD5}), to keep the dependence of the torsion on the first derivatives of $\phi$ and to avoid the dependence of the curvature on the second derivatives. 

It is easy to reintroduce the variables $\omega^{[ik]}$ used in ref.\ \cite{CSTVZ}.  By means of eq. (\ref{Scaled}), we obtain the Lagrangian
\begin{equation} \label{GBDLag2}
\lambda^G =  2^{-2} \phi^2 \epsilon_{ikjl} (d(\phi \omega^{[ik]}) +\phi^2  g_{mn} \,  \omega^{[im]} \wedge  \omega^{[nk]}) \wedge  \omega^j \wedge  \omega^l + \lambda^+ ,
\end{equation}
and the field equations
\begin{eqnarray}  \label{HorDiff4}
&F_{p[jl]}^{[ik]} = \beta_p (\delta^i_j \delta^k_l - \delta^i_l \delta^k_j), \qquad 
F_{[jl][mn]}^{[ik]} = \phi \hat F_{[jl][mn]}^{[ik]},  \nonumber \\
&\phi^3 (- F_{jk}^{[ik]} + 2^{-1} \delta^i_j F_{lk}^{[lk]})  = T^{\phi i}_j + T^i_j, 
\end{eqnarray}
\begin{eqnarray} \label{VerDiff4}
&F_{[jk]l}^i = \phi \hat F_{[jk]l}^i, \qquad F_{[jk][mn]}^i = 0,  \nonumber \\
&\phi^3 (F_{jk}^i + \delta^i_j F_{kl}^l - \delta^i_k F_{jl}^l) 
+ 2 \phi^3 (\beta_j \delta^i_k- \beta_k\delta^i_j) = T^i_{[jk]}.
\end{eqnarray} 

From these equations we  obtain eq.\ (\ref{PhiDef}) and the formulas 
\begin{equation} \label{Beta3}
\beta_i  = 3^{-1} F_{i k}^{k} + 6^{-1} \phi^{-3} T_{[ik]}^{k} = 12^{-1} F_{i[jk]}^{[jk]},
\end{equation}  
that show the geometric nature of the fields $\phi$ and $\beta_i$.

\section{The fully augmented covariance group}
\label{Groups} 
Perhaps, the first example of augmented Lorentz covariance was provided by Born's reciprocity theory \cite{Born1,Born2}. This theory has covariance transformations that mix the spacetime coordinates $x^i$ and the components $P_i$ of energy-momentum.  More recently, similar ideas have been discussed \cite{Schuller,SP,Low1,Low2}.

It is interesting to note the analogies and the differences between Born's reciprocity principle and the equity principle proposed in ref.\ \cite{Lurcat}. The spacetime coordinates are compared with the components of 4-momentum in the first case and with the components of the 4-velocity in the second case.

Another example of augmented Lorentz group is given by the metric-affine gravitational theory
\cite{BH,HMMN}, in which the linear group $GL(4, \mathbf{R})$  (a subgroup of the  affine gauge group)  acts on the tangent spaces of the spacetime manifold.  The group is the same that we consider in the following, but its action on the physical fields is rather different.

After the general discussion of ref.\ \cite{Toller4}, a more definite choice of the augmented covariance group $\mathcal{C}$, containing the orthochronous Lorentz group and provided with an irreducible representation acting on $\mathcal{T}$, was proposed in refs.\ \cite{Toller5,Toller9}. 

This proposal starts from an operational analysis of the kinematical concepts given in ref. \cite{Toller2}. In this analysis  the local Lorentz frames (tetrads) are defined by some material objects that necessarily are involved in the description of any physical operation. A central role is played by the concept of ``feasible transformation'',  that describes the physical operations necessary to build a new reference frame starting from a pre-existent one.  
 
The infinitesimal feasible transformations are represented by vector fields belonging to a wedge $\mathcal{T}^+ \subset \mathcal{T}$. It is a wedge, namely it is dilatation invariant and convex, because by performing two successive feasible transformations one obtains another feasible transformation. 

It is  natural to assume that $\mathcal{T}^+$ is invariant under the covariance group. We also assume that it contains the generator $A_0$ of the positive time translations and that it does not contain $- A_0$. The mathematical treatment  is simpler if we  assume that  $\mathcal{T}^+$ is closed, namely it coincides with its closure  $\overline\mathcal{T}^+$.  This means that we consider as feasible also transformations that can be approximated with arbitrary precision by really feasible transformations.
 
The vector subspace
\begin{equation} \label{Reversible}
\mathcal{T}_R = \overline\mathcal{T}^+ \cap - \overline\mathcal{T}^+ \subset \mathcal{T}
\end{equation}
contains the ``reversible'' infinitesiomal transformations. Since it does not contain $A_0$ and it is invariant under the augmented covariance group, the equity principle requires that it has dimension zero. This means that the wedge $\overline\mathcal{T}^+$ is actually a cone. 

Also the linear subspace generated by $\mathcal{T}^+$ is invariant and it must coincide with $\mathcal{T}$. This means that $\overline\mathcal{T}^+$  has a non empty interior.

In a theory that  admits arbitrarily large accelerations,  the infinitesimal Lorentz transformations are reversible, namely $\mathcal{T}_V \subset \mathcal{T}_R$,  the wedge $\mathcal{T}^+$ is not a cone and  the  equity principle is not satisfied.

Note the analogy between  $\overline\mathcal{T}^+$ and  the closed future cone $\overline\mathcal{V}^+$, contained in the Minkowski spacetime or in a tangent space of the curved spacetime $\mathcal{M}$.   $\overline\mathcal{V}^+$ describes an upper bound to the velocity.

In agreement with the equity principle, the cone $\overline\mathcal{T}^+$ describes, besides an upper bound to the velocity, also upper bounds to the angular velocity and to the acceleration. The maximal acceleration has been introduced and discussed in refs.\ \cite{Caianiello,CDFMV,Brandt1,Scarpetta,Brandt2,Papini,Toller16}.  The idea of minimal length is strictly connected with the idea of maximal acceleration, but, in our opinion, the latter is more clearly defined in terms of classical concepts.

The arguments that lead to a specific choice of $\overline\mathcal{T}^+$ consist of double interplay (like a ping-pong game) between symmetry groups and closed cones, summarized in the following scheme:
\begin{eqnarray} \label{PingPong}
O(3) & & \nonumber \\
& \searrow & \nonumber \\
& & \overline\mathcal{V}^+ \nonumber \\
& \swarrow & \nonumber \\
O(1,3)^{\uparrow}& & \nonumber \\
& \searrow & \nonumber  \\
& & \overline\mathcal{T}^+ \nonumber  \\
& \swarrow & \nonumber \\
GL(4, \mathbf{R}). & & 
\end{eqnarray} 

It is easy to show that the symmetry with respect to the rotation group $O(3)$, together with the choice of a maximal valocity $c$, determines the closed future cone $\overline\mathcal{V}^+$. Disregarding the spacetime dilatations, the symmetry group of this cone is the orthochronous Lorentz group $O(1,3)^{\uparrow}$. 

It is not so evident, but it is proven in refs.\ \cite{Toller5,Toller9}, that the Lorentz symmetry, together with the choice of a maximal acceleration $c^2/\tilde\ell$ (or of a fundamental length $\tilde\ell$), determines the cone $\overline\mathcal{T}^+$. The symmetry group of this cone is $GL(4, \mathbf{R})$. We call it the ``fully augmented'' (geometric) covariance group, because the actual geometric covariance group $\mathcal{C}$, that respects the cone $\overline\mathcal{T}^+$,  has to be chosen among its subgroups that contain $O(1,3)^{\uparrow}$.

In  general, the constant $\ell$, that appears in the definition of the cone $\overline\mathcal{T}^+$ and of its symmetry group, has dimension
\begin{equation} \label{Dimension}
[\ell] = [A_i]^{-1} [A_{[ik]}] = [\omega^i] [\omega^{[ik]}]^{-1}.
\end{equation} 
In theories with constant gravitational coupling, as GR and the EC theory, $A_{[ik]}$ generates directly the Lorentz transformations and therefore is dimensionless (since we are using the convention $c = 1$, time and length have the same dimension $[L]$). Since $[A_i] = [L^{-1}]$, it follows that $\ell = \tilde\ell$ is a fundamental length.
  
As we have explained in Section \ref{Revisited} (see eq.\ (\ref{Scaled})), in the GBD theory covariant under total dilatations, the fields $A_{[ik]}$ are proportional to the scalar field $\phi$, related to the variable gravitational coupling as in eq.\ (\ref{GBD}). In this case we have 
\begin{equation} \label{Dimension1}
[A_{[ik]}] = [\phi] = [L^{-1/2} M^{1/2}], \qquad [\ell] = [L^{1/2} M^{1/2}],
\end{equation}
namely $\ell^2$ is an action. 

The minimal length, that limits the application of the usual space-time concepts, is given by
\begin{equation} \label{Minimal}
\tilde \ell = \phi^{-1} \ell
\end{equation}
and, if $G$ is variable, it depends on the frame $s \in \mathcal{S}$.  Several authors \cite{Hossenfelder2,Ferretti,Garay} have suggested that it should be equal, up to a factor of the order of one, to the Planck length
\begin{equation} \label{Planck}
\tilde \ell \approx \ell_P = (\hbar G)^{1/2}.
\end{equation}
In this case, we have $8 \pi \ell^2 \approx \hbar$. Note, however, that a considerably larger value of $\tilde \ell$ is compatible with the experimental observations.

The ``geometric fundamental action'' $\ell^2$ and the ``quantum of action'' $\hbar$ appear independently in two different parts of theoretical physics. Another different fundamental constant with the dimension of an action (if $c = 1$) is the square $e^2$ of the elementary charge.

The ratio $e^2 / \hbar \approx 1/137$ is known numerically with very high precision but no theoretical explanation of its experimental value is known.  For our theoretical speculations, it is preferable to replace $e$ with the coupling constant $g$ of a grand unified theory, calculated at the grand unification energy scale \cite{Weinberg2}.  

The calculation of  $g$, starting from the known values of $e$ and of the measured strengths of the other particle interactions, is a task of a quantum field  theory of particle interactions and it depends on the adopted model. For instance, ref.\ \cite{AM} suggests the value $g^2 = \hbar / 36.4$ at a grand unification energy of about $10^{16} GeV$.

The three fundamental actions can be represented as the vertices of a ``triangle of actions'':
\begin{eqnarray}  \label{Triangle}
& \ell^2  & \nonumber \\
&\nearrow  \qquad  \searrow &  \nonumber \\
\hbar & \longleftarrow & g^2  
\end{eqnarray}
The arrows represent ratios between the constants and involve different theoretical concepts.  The left arrow indicates a problem of quantum gravity, the right arrow concerns the ideas of supercovariance or supergravity and, as we have already observed, the lower arrow can be treated successfully by means of the ideas of experimental and theoretical elementary particle physics. 

Also the  constant $\hat \Lambda $,  that appears in eq.\ (\ref{GBDLagPlus})  and is connected with the cosmological constant, has  the dimension of an action, but its ratio with the other three fundamental actions is a very large number, that can hardly be explained by any theory.

Since we have only a rather large experimental upper bound for $\ell^2$, the ratios associated to the left an right arrows are very badly known.  We hope that our efforts to clarify the group-theoretical meaning of the action $\ell^2$ can provide some help in the treatment of these difficult problems.  Some ideas suggested by the triangle (\ref{Triangle}) are shortly presented in  Section \ref{Conclusions}.

\section{The spinor formalism}
\label{Spinor} 

In order to find an explicit definition of $\overline\mathcal{T}^+$ with the required properties, it is convenient to use the Dirac spinor formalism. We use a Majorana basis, in which the matrices $\gamma_i$ are real and satisfy the equation
\begin{equation}
\gamma_i \gamma_k + \gamma_k \gamma_i = 2 g_{ik}.
\end{equation}
We also consider the real antisymmetric matrix $C$ with the properties
\begin{equation} 
\gamma_i^T = \breve C \gamma_i C, \qquad \breve C = - C^{-1},  \qquad - 2^{-3} \epsilon^{ABCD} C_{AB} C_{CD} = 1,
\end{equation}
where $\gamma_i^T$ is the transposed matrix and $\epsilon^{ABCD}$ is the totally antisymmetric spinor with $\epsilon^{1234} = 1$. Given an explicit representation of the $\gamma$ matrices, these equations determine the matrix $C$ up to its sign and we choose it in such a way that the symmetric matrix $\gamma^0 C$ is positive definite.

We write a vector $A \in \mathcal{T}$ in the form 
\begin{equation} \label{TCoordinates}
A =  a^{\alpha} A_{\alpha} = a^i A_i + 2^{-1} a^{[ik]} A_{[ik]}
\end{equation} 
and we consider the real symmetric $4 \times 4$ matrix
\begin{equation} \label{MatrixB}
\mathrm{a} = a^{\alpha} \breve \Xi_{\alpha}.
\end{equation} 
where
\begin{equation} \label{Gamma1}
\breve \Xi_i = - \breve C \gamma_i, \qquad
\breve \Xi_{[ik]} = - 2^{-1} \ell  \breve C (\gamma_i \gamma_k - \gamma_k \gamma_i).
\end{equation}

We also define the real symmetric matrices
\begin{equation} \label{Gamma2}
\Xi^i =  \gamma^i C, \qquad
\Xi^{[ik]} = - (2 \ell)^{-1}  (\gamma^{i} \gamma^{k} - \gamma^{k} \gamma^{i}) C,
\end{equation}
and from the property 
\begin{equation}
2^{-2} \mathrm{Tr} (\Xi^{\alpha} \breve\Xi_{\beta}) = \delta^{\alpha}_{\beta},
\end{equation}
we obtain the inverse formula
\begin{equation} \label{Inver}
a^{\alpha} = 2^{-2} \mathrm{Tr}  (\Xi^{\alpha} \mathrm{a}).
\end{equation}

The cone $\overline\mathcal{T}^+$ is defined by requiring that the matrix $\mathrm{a}$ is positive semidefinite, namely that
\begin{equation} 
\psi^T \mathrm{a} \psi \geq 0
\end{equation}
for any choice of the real spinor $\psi$. 

It is clear that $\overline\mathcal{T}^+$ is invariant under the transformations
\begin{equation} \label{SymmTrans}
\mathrm{a} \to \mathrm{m}^T \mathrm{a} \mathrm{m}, \qquad \mathrm{m} \in GL(4, \mathrm{R}).
\end{equation}
This transformation property means that the matrix $\mathrm{a}$ with elements $a^{AB} = a^{BA}$ represents a contravariant symmetric $GL(4, \mathrm{R})$ spinor.  

The multiples of the unit matrix form a subgroup of $GL(4, \mathrm{R})$ that describes the total dilatations.  The matrices with $\det \mathrm{m} = 1$ form the connected subgroup $SL(4, \mathrm{R})$.  It contains a subgroup isomorphic to $SL(2, \mathrm{C})$, locally isomorphic to the proper orthochronous Lorentz group. Its infinitesimal transformations are given by the matrices
\begin{equation} \label{SigmaIK}
\Sigma_{ik} = 2^{-2}(\gamma_i \gamma_k - \gamma_k \gamma_i)
\end{equation}
acting on the covariant spinor indices. The matrices $\pm \gamma_0 \in SL(4, \mathrm{R})$ describe the space reflection (parity), that changes the sign of $a^1, a^2, a^3, a^{[01]}, a^{[02]}, a^{[03]}$. We see that, as we  required in the preceding Section \ref{Groups}, the cone $\overline\mathcal{T}^+$ is invariant under the orthochronous Lorentz group.

By means of the matrices (\ref{Gamma1}) and (\ref{Gamma2}), we can write various physical quantities in the spinor formalism, for instance we define the symmetric spinors
\begin{equation} \label{SymmSpin}
A_{AB} = \Xi^{\alpha}_{AB} A_{\alpha},  \qquad
A_{\alpha} = 2^{-2} \breve \Xi_{\alpha}^{AB} A_{AB},
\end{equation} 
\begin{equation} \label{SymmSpin2}
\omega^{AB} = \breve\Xi_{\alpha}^{AB} \omega^{\alpha}, \qquad
\omega^{\alpha} = 2^{-2} \Xi^{\alpha}_{AB} \omega^{AB}.
\end{equation} 

We shall also use covariant and contravariant antisymmetric spinors of the kind $f_{AB} = - f_{BA}$ and $h^{AB} = - h^{BA}$. They can be represented by antisymmetric matrices $\mathrm{f}$  and $\mathrm{h}$ that can always be written in the form
\begin{equation} \label{MatrixF}
\mathrm{f} = f_u \Theta^u, \qquad
f_u = - 2^{-2} \mathrm{Tr}  (\breve\Theta_u \mathrm{f}),
\end{equation}
\begin{equation} \label{MatrixH}
\mathrm{h} = h^u \breve\Theta_u, \qquad
h^u = - 2^{-2} \mathrm{Tr} (\Theta^u \mathrm{h}).
\end{equation}

We have introduced the antisymmetric real matrices $\breve\Theta^u_{AB}$ and $\Theta_u^{AB}$ given by
\begin{eqnarray} \label{Theta}
&\Theta^i = - \gamma^i \gamma^5 C, \qquad
\Theta^4 = C, \qquad
\Theta^5 = - \gamma^5 C = - G,
\end{eqnarray} 
\begin{equation} \label{ThetaBreve}
\breve\Theta_i =  \breve C \gamma_i \gamma_5, \qquad
\breve\Theta_4 = \breve C, \qquad  \breve\Theta_5 = - \breve C \gamma_5 = - \breve G.
\end{equation} 
\begin{equation} \label{Gamma5}
\gamma_5 = - \gamma^5 = \gamma_0 \gamma_1 \gamma_2 \gamma_3, \qquad
\gamma_5^T = - \breve C \gamma_5 C.
\end{equation}
In the following, we use the identity
\begin{equation} \label{GammaSigma}
C (\gamma^i \Sigma_{jk} + \Sigma_{jk} \gamma^i) = \epsilon_{jk}{}^{il} \Theta_l.
\end{equation}

Starting from the equations we have written above, one can prove the identity
\begin{equation} \label{Useful2}
2^{-3} \epsilon^{ABCD} f_{AB} f_{CD} = g^{uv} f_u  f_v, \qquad g^{44} = g^{55} = - 1.
\end{equation}
In the present Section, the indices $u, v, w, x, y, z$ take the values $0,\ldots, 5$, while the indices $i, k, j, l, m, n$ take the values $0,\ldots, 3$. The left hand side of this equation is invariant under the action of $SL(4, \mathrm{R})$ on the antisymmetric spinors. 

We see that, when this action is expressed in terms of the components $f_u$, we obtain a transformation of the pseudo-orthogonal group $SO(3, 3)$. Actually, one defines in this way an isomorphism between $SL(4, \mathrm{R})$ and a double covering of the identity component $SO(3, 3)^c$ of $SO(3, 3)$. 

The identity connected component $GL(4, \mathrm{R})^c$ contains the matrices with positive determinant and is generated by $SL(4, \mathrm{R})$ and the total dilatations.  The second connected component contains the matrices with negative determinant. They have a rather obscure physical meaning and we do not consider them in the following.  $GL(4, \mathrm{R})^c$ is locally isomorphic  to the product of  $SO(3, 3)^c$  and the group $\mathrm{R}_{td}^+$ of the total dilatations.

Note that the orthochronous Lorentz group $O(1,3)^{\uparrow}$ can be considered as a subgroup of $SO(3, 3)^c$. The proper transformations act in the usual way on $f_0,\ldots, f_3$, while the space reflection is represented by a rotation that changes the signs of the components $f_0$ and $f_5$. 

We indicate by $X_{uv}$ the operators that form a basis of the Lie algebra $o(3, 3)$. The infinitesimal transformations of a 6-vector  $\xi_w$ are given by
\begin{equation} \label{XVWF}
X_{uv} \xi_w = g_{wu} \xi_v - g_{wv} \xi_u.
\end{equation}

Using a matrix notation, their actions on a covariant spinor field $\Psi$ can be written as
\begin{equation} \label{SpinorTrans}
X_{uv} \Psi = \Sigma_{uv} \Psi,
\end{equation}
where the matrices $\Sigma_{uv}$ are given by eq.\ (\ref{SigmaIK}) and
\begin{eqnarray} \label{SigmaUV}
&\Sigma_{i4} = - \Sigma_{4i} = 2^{-1} \gamma_i \gamma_5, \qquad
\Sigma_{i5} = - \Sigma_{5i} = 2^{-1} \gamma_i,& \nonumber \\
&\Sigma_{45} = - \Sigma_{54} = 2^{-1} \gamma_5.&
\end{eqnarray}

They satisfy the commutation relations
\begin{equation} \label{SigmaCommu}
[\Sigma_{uv}, \Sigma_{xy}]  
= g_{vx} \Sigma_{uy} - g_{ux} \Sigma_{vy} 
- g_{vy} \Sigma_{ux} + g_{uy} \Sigma_{vx}.
\end{equation}

The covariance  property of the matrices $\breve \Theta_u$ is given by
\begin{equation} \label{ThetaSigma}
\breve \Theta_u \Sigma_{vw} +  \Sigma_{vw}^T \breve \Theta_u = 
g_{uv} \breve \Theta_w - g_{uw} \breve  \Theta_v.
\end{equation}

Also the symmetric covariant or contravariant spinors, for instance the quantities (\ref{SymmSpin}) and (\ref{SymmSpin2}), are equivalent to 6-dimensional tensors.  Since they must have 10 independent components, there is no choice: they are fully antisymmetric self-dual or anti-self-dual tensor with three indices.  We do not use this rather heavy formalism in the present article.

\section{A partially augmented covariance group}
\label{Partially} 
In his construction of Special Relativity, Einstein  modified the equations of classical mechanics in a Lorentz covariant way (Maxwell's theory was already Lorentz covariant). In a similar way, one could try to modify the relativistic equations to obtain equations with augmented covariance. This is a difficult problem and, 35 years after the proposal of the $GL(4, \mathrm{R})$ augmented covariance group \cite{Toller5}, no satisfactory solution has been found.

In particular, an acceptable Lagrangian theory of gravitation with  fully augmented covariance  is not yet available. A first problem arises because the space reflection, represented by $\pm \gamma_0$, belongs to the connected subgroup $SL(4, \mathrm{R})$. It follows that a pseudoscalar fully covariant Lagrangian form cannot exist.

The Lagrangian forms (\ref{NR}) and (\ref{GBDLag}) contain the pseudotensor $\epsilon_{ijkl}$ and  are pseudoscalar. As a consequence, the conserved 3-form $\tau_i$, that describe the energy-momentum density, are  Lorentz pseudo 4-vectors.   This is a necessary feature if one wants to compensate the change of orientation of the integration surfaces $S$ and $\Sigma$  appearing in eqs.\ (\ref{ActionPrinciple}) and (\ref{TenMomentum}) when one performs a space reflection.

In a fully covariant theory, a similar compensation has to be obtained by introducing new dynamical variables, that can generate a parity doubling of the vacuum  states, namely a spontaneous breaking  of the space reflection covariance. This can be considered  as an interesting feature in view of a connection with the chiral properties of the weak interactions.  These problems will be examined in detail in a forthcoming article.

In the following we describe a simpler model based on a partially augmented covariance group $\mathcal{C}$, that does not contain the space reflection in its connected component of the unit $\mathcal{C}^c$,  but  is sufficient to implement the equity principle, namely it has an irreducible 10-dimensional representation acting on $\mathcal{T}$.  The space reflection is contained in a second connected component and the action integral is invariant under space reflection,  provided that the Lagrangian is pseudoscalar. 

The analysis of ref.\ \cite{Toller4} suggests to use one of the many anti-de Sitter subgroups  $SO(2,3)^c$ contained in $SO(3,3)^c$.  It is locally isomorphic to a symplectic subgroup $Sp(4, \mathrm{R)} \subset SL(4, \mathrm{R})$. The total dilatations can be treated separately.

In order to specify the choice of $\mathcal{C}^c$, we have to choose a 6-vector $f_u$ invariant under the relevant $SO(2,3)^c$ subgroup and the corresponding antisymmetric spinor $f_{AB}$ (see eq.\ (\ref{MatrixF})) used in the definition of $Sp(4, \mathrm{R)}$.  

Since $\mathcal{C}^c$ contains the proper orthochronous Lorentz group, we must have $f_i = 0$ namely $f = (0, 0, 0, 0, f_4,  f_5)$.  Since  $f_u$ cannot be invariant under space reflection, we must also have $f_5 \neq 0$.  If we assume that $\mathcal{C}$  contains the space reflection in a second connected component,  $\mathcal{C}^c$ must also contain the anti-de Sitter subgroup that does not affect $f' = (0,  0, 0, 0, f_4,  - f_5)$,  and, unless $f_4 = 0$, the two anti-de Sitter subgroups generate the whole group $SO(3,3)^c$. 

We have seen that the only possible choice is
\begin{equation} 
f_u = \delta^5_u, \qquad  f_{AB} = G_{AB} = (\gamma^5 C)_{AB}.
\end{equation}
We indicate by $SO(2,3)_A^c$ the corresponding  anti-de Sitter subgroup, that does not act on the index $u = 5$ and cannot contain the space reflection. The subscript ``A''  stands for ``Axial vector'',  to indicate the behavior under Lorentz transformations of four elements $X_{i4}$ of the Lie algebra. We also indicate by $Sp(4, \mathrm{R})_A$ the corresponding symplectic group defined by the antisymmetric spinor $G_{AB}$.   

Many theoretical investigations deal with the group  $SO(2,3)_V^c$,  where ``V''  stands for ``Vector'', because the additional four elements of the Lie algebra transform as ordinary Lorentz 4-vectors, as it happens for the Poincar\' e group.  The corresponding symplectic group $Sp(4, \mathrm{R})_V$ is defined by the antisymmetric spinor $C_{AB}$ and contains $\pm \gamma_0$.   

The invariant spinor $G_{AB}$ and its inverse $- \breve G^{AB}$ can be used to lower and to rise the spinor indices. This operation is not allowed in a theory with fully augmented covariance.  

In conclusion, we have seen that, in the spinor formalism, the partially augmented covariance group $\mathcal{C}$ is generated by $Sp(4, \mathrm{R})_A$, the total dilatations and the space reflection $\pm \gamma_0$ that generates a second connected component. 

Another motivation for considering the partially augmented covariance group $Sp(4, \mathrm{R})_A$ appears when one tries to introduce a supercovariance group, that  mixes Bosonic and Fermionic indices.  The connection between supercovariance and maximal acceleration has been briefly discussed several years ago in ref.\ \cite{Toller10}.  The introduction of  an  augmented covariance group requires a careful reexamination  of many ideas concerning supercovariance.

Note that now we give a different meaning to the words supercovariance and supersymmetry or supergravity \cite{Weinberg3,Ferrara}.  According to the definition we have adopted in Section \ref{Introduction}, a covariance transformation has a global nature and acts only on the field indices, while supersymmetry transformations have a more general nature.

Abe and Nakanishi in a series of articles (see for instance \cite{Nakanishi1,AN6}), have proposed a supercovariant extension of the Lorentz covariance under $SL(2, \mathrm{C}) = Sp(2, \mathrm{C})$, that they call ``New Local Supersymmetry'' (NLS).  They suggest the supercovariance group $OSp(N|2; \mathrm{C})$ \cite{FSS,DeWitt}, that contains, besides  $Sp(2, \mathrm{C})$, the non compact internal covariance group $SO(N,\mathrm{C})$. 

A natural supercovariant extension of the fully augmented covariance group is the super Lie group  $SL(N| 4; \mathrm{R})$.  This group contains, besides  $SL(4, \mathrm{R})$ and (for $N \neq 4$) a one-parameter group of ``superdilatations'',  a non compact real internal symmetry group  $SL(N, \mathrm{R})$, that has no finite dimensional unitary representation that could describe finite particle multiplets.

A possible way to avoid this problem is to consider the orthosymplectic subgroup $OSp(N | 4; \mathrm{R})$ \cite{FSS,DeWitt}, that contains the group $Sp(4, \mathrm{R})_A$ considered in the present article and the compact real internal covariance group $SO(N, \mathrm{R})$.  The possible relation between the latter group and a realistic (but broken)  grand unification group \cite{Weinberg2} deserves further investigations.  

A third argument in favor of the partially augmented covariance group $SO(2,3)_A^c$ appears in the treatment of the test particles given in refs.\ \cite{Kunzle,Vanzo,Toller6,TollerVaia}. The Dixon condition \cite{Dixon} 
\begin{equation} \label{Dixon}
P_{[ik]} P^k = 0,
\end{equation}
that characterizes the Lorentz frames in which the particle trajectory crosses the origin, using the 5-dimensional tensor formalism introduced below, can be written in the form
\begin{equation} \label{Dixon2}
\epsilon^{uvwxy} P_{ vw}  P_{ xy} = 0,
\end{equation}
that is covariant under $SO(2, 3)_A^c$, but not under  under $SO(2, 3)_V^c$.

Some authors, see for instance refs.\ \cite{MDM,West,Fre,SW}, have proposed gravitational theories symmetric under an anti-de Sitter group $SO(2, 3)^c$ or its double covering $Sp(4, \mathrm{R})$. The proposals we have seen look rather different from the theory we are discussing and have different motivations. In particular, they are local theories based on a spacetime manifold and do not satisfy the equity principle.   In general, they deal with the group we have called $SO(2, 3)_V^c$.

It is convenient to formulate the theories with partially augmented covariance in terms of 5-dimensional tensors, with indices $u,v,\ldots, z$ that take the values $0,\ldots, 4$.   $SO(2, 3)_A^c$ operates in the usual way and the space reflection, that does not belong to it, introduces a minus sign for each index that takes the value 0, an operation with determinant $-1$, when only one index is considered. If one deals with pseudotensors, the space reflection introduces one more minus sign. For instance, if $f_u$ is a pseudo 5-vector, the first four components behave as the components of a Lorentz 4-vector.

The quantities with ten independent components that transform irreducibly, in particular  $A_{\alpha}$ and $\omega^{\alpha}$, can only be represented by antisymmetric pseudotensors with two indices.  In fact, from the tensorial quantities $A_{uv}$ and $\omega^{uv}$  we can define the symmetric spinors
\begin{equation}  \label{SymmSpin3}
A_{AD} = 2^{-1} A_{uv} \Theta^u_{AB} \breve G^{BC} \Theta^v_{CD}, 
\end{equation}
\begin{equation} \label{SymmSpin4}
\omega^{AD} = 2^{-1}  \omega^{uv} \breve\Theta_u^{AB} G_{BC} \breve\Theta_v^{CD}.
\end{equation}
Since the right hand sides contain $G$ or $\breve G$, these formulas are covariant only under $Sp(4, \mathrm{R)}_A^c$ and also under space reflection if the quantities $A_{uv}$ and $\omega^{uv}$ transform as pseudotensors in 5 dimensions.

From eqs.\ (\ref{SymmSpin}) and (\ref{SymmSpin2}), we have
\begin{equation}
A_{\alpha} =  2^{-3} A_{uv} \mathrm{Tr} (\breve \Xi_{\alpha} \Theta^u \breve G \Theta^v), \qquad
\omega^{\alpha} = 2^{-3} \omega^{uv} \mathrm{Tr} (\Xi^{\alpha} \breve\Theta_u G \breve\Theta_v)
\end{equation}
and, computing the traces, we find the following relations between the Lorentz tensors and the 5-dimensional tensors:
\begin{equation} \label{ALor}
A_{[ik]} = - 2^{-1} \ell \epsilon_{ik}{}^{jl} A_{jl}, \qquad
A_i =  A_{i4},
\end{equation}
\begin{equation}  \label{OmegaLor}
\omega^{[ik]} =  (2 \ell)^{-1} \epsilon^{ik}{}_{jl} \, \omega^{jl}, \qquad 
\omega^i = \, \omega^{i4}.
\end{equation}

Note that these formulas behave correctly with respect to the space reflection.  Similar formulas can be applied to other quantities, but not to the transformation matrices: the notations $\Sigma_{[ik]}$,  $\Sigma_i$ and $\Sigma_{\alpha}$ are misleading and should be avoided.

We shall also use the formula
\begin{equation} \label{XI4Omega}
X_{uv} \omega^{xy} = (\delta_u^x g_{vz}  - \delta_v^x g_{uz}) \,\omega^{zy}
+  (\delta_u^y g_{vz}  - \delta_v^y g_{uz}) \,\omega^{xz}.
\end{equation}
From eq.\ (\ref{ThetaSigma})  we obtain the useful identity
\begin{equation} \label{GSigma}
\breve G \Sigma_{vw} =  (\breve G\Sigma_{vw})^T , \qquad v, w < 5,
\end{equation} 
that confirms the invariance  property of the matrix $\breve G =\breve \Theta_5$  under $Sp(4, \mathrm{R)}$.

\section{The manifold of the vacuum states}
\label{Vacuum} 
Before concentrating our attention on the details of a partially covariant theory, we consider the properties of its vacuum states.  They are defined as the states in which all the physically relevant  fields defined on $\mathcal{S}$  are constant.  

In particular, the structure coefficients, that have to satisfy eq.\ (\ref{Jacobi}), are the structure constants of a Lie algebra and, if we choose a point representing the unit element,  $\mathcal{S}$ becomes (at least locally)  a Lie group. We disregard more complicated global topological structures.

We assume that there is a vacuum state, that we call the ``standard vacuum'' in which the structure coefficients are  invariant under the orthochronous Lorentz group $O(3,1)^{\uparrow}$.  We also assume  that all the other vacuum states can be obtained from it by means of a transformation of  the covariance group $\mathcal{C}$.

In the standard vacuum states we shall consider, $\mathcal{S}$ is isomorphic (as a manifold with absolute parallelism) to the Poincar\'e or the de Sitter group and the non vanishing structure coefficients are given by eqs.\ (\ref{StructConst1}), (\ref{StructConst2}) and  (\ref{StructConst3}) with a suitable non negative value of the constant $K$.  If we also consider the group $\mathbf{R}^+_{td}$ of the total dilatations, we can multiply all the structure constants by  a constant  $\phi = \hat\phi$.

We have assumed that the vacuum states form a set on which the covariance group $\mathcal{C}$ acts transitively and the stabilizer of a point representing a standard vacuum is  $O(3,1)^{\uparrow}$. It follows that  the vacuum states form an  homogeneous space isomorphic to $\mathcal{C} / O(3,1)^{\uparrow}$.  

We also consider, in the space of the 5-vectors, the open set  defined by
\begin{equation} \label{Aug}
 - \xi_u \xi^u = \phi^2, \qquad  \phi  > 0.
\end{equation}
The covariance group $\mathcal{C}$ acts transitively on it and the stabilizer of the element $\hat \xi_u = \hat \phi \delta_u^4$  is $O(3,1)^{\uparrow}$. It follows that eq.\ (\ref{Aug}) defines an homogeneous space isomorphic to $\mathcal{C} / O(3,1)^{\uparrow}$ and to the space of the vacuum states. The vector $\hat \xi_u$ corresponds to the standard vacuum.
 
One can write the structure coefficient of the vacuum states as uniquely defined functions of  $\xi_u$ by means of an explicit formula that we do not need in the following.  There are many explicit inverse formulas that give $\xi_u$ as a function of the vacuum structure coefficients. One can extend one of these formulas to all the field configurations of the theory and consider $\xi_u$ as a variable geometric field. An example is given in ref.\ \cite{Toller4}.  In the next Section \ref{Model} we adopt a different point of view and the connection of $\xi_u$ with the structure coefficients follows from the field equations.

Since $\xi_u$ appears in connection with a spontaneously broken augmented covariance group, we call it the ``augmentonic'' field. The equation (\ref{Aug}) gives the dilatonic field $\phi$ as a function of the augmentonic field. It can be considered as a  particular augmentonic field associated with the introduction of the (spontaneously broken) covariance under total dilatations. Following the example given by the GBD theory described in Section \ref{Revisited}, we use the augmentonic  field as an instrument for the construction of more general theories with augmented covariance.  

One should also study the analogy between the five components of the augmentonic field and the massless  Goldstone fields \cite{Weinberg6, Strocchi}, that, in a local theory, would appear  in connection  with the five  spontaneously broken infinitesimal generators of the covariance group.  The problem is that,  as far as we know,  no general Goldstone theorem has been proved for theories of the kind we are considering.  In any case, we shall see in the next Section \ref{Conservation} that the augmentonic fields satisfy zero mass equations.

The moment mapping (see Section \ref{General}) maps the vacuum states into elements of the dual $L(C)^* = so(2, 3)^*$ of the covariance Lie algebra represented by antisymmetric 5-tensor with two indices.  This mapping preserves the covariance group and, since the vacuum states form an orbit, they are mapped into a  ``coadjoint orbit'' of $L(C)^*$.  There is a Lorentz invariant vacuum state, but the only coadjoint orbit that contains a Lorentz invariant element is the trivial orbit $\{ 0 \}$.  In other words, all the conserved quantities $Q_{uv}$ vanish in the vacuum states.

It is known \cite{Woodhouse,Souriau} that the invariant symplectic form defined on the whole (infinite dimensional) space of the states, when restricted to an orbit of the covariance group,  becomes an invariant presymplectic form, which is the pullback of a symplctic form naturally defined on the corresponding coadjoint orbit.  In the case we are considering, the presymplectic form defined in this way on the orbit of the vacuum states vanishes.

This negative conclusion is not valid any more if one considers a theory with fully augmented covariance, since in this case $L(C)^*$ contains non vanishing Lorentz invariant elements. In particular  $Q_{45}$ does not necessarily vanish in the standard vacuum.  This remark suggests the introduction of some quantum properties of the vacuum states (see Section \ref{Conclusions}).

\section{A model with partially augmented covariance}
\label{Model} 

Now we have to find a Lagrangian invariant under  $SO(2,3)^c_A$ and the total dilatations and odd under the space reflection. A gravitational Lagrangian with this property was given in ref.\ \cite{Toller4}, but it is not satisfactory. The problem is that the augmentonic field is ``frozen'', namely it is constant as a consequence of some field equations.

The  Lagrangian of ref.\ \cite{Toller4}, as the Lagrangian of the GBD theory proposed in ref.\ \cite{CSTVZ},  depends only on the forms $\omega^{\alpha}$ and on the structure coefficients. It also contains a field $\xi_u$ transforming as a 5-vector, but this field is defined from the beginning as a known function of the structure coefficients. 

More interesting Lagrangians can be obtained by means of a generalization of the ideas used in Section \ref{Revisited} for a reformulation of the GBD theory, namely by considering the augmentonic field  $\xi_u$ as an independent dynamical variable. Its dependence on the structure coefficients, that implies its geometric nature, will appear as a consequence of the field equations.

We work in a first order formalism and,  to obtain  second order equations for the field $\xi_u$, we have to introduce some other independent ``auxiliary'' fields.  In the following we adopt a minimal choice of these new fields,  namely we introduce, besides the 5-vector $\xi_u$,  a 5-pseudovector $\beta_u$, that generalizes the 4-vector  $\beta_i$ that appears in the GBD theory of Section \ref{Revisited}. This field too has a geometric nature, since the field equations (\ref{TanBeta4}) show that it depends on the derivatives of $\xi_u$ and the eqs.\ (\ref{BetaGeom}) give it directly as a function of the structure coefficients.

We choose the geometric Lagrangian in such a way that, for
\begin{equation} \label{Special2}
\xi_i = 0, \qquad \xi_4 = \phi
\end{equation} 
and $\beta_4 = 0$ it coincides with the GBD Lagrangian  (\ref{GBDLag2}). In particular, we put,  in the 5-dimensional tensor formalism,
\begin{eqnarray}  \label{LagAug}
&\lambda^G = (2 \ell \phi)^{-1}  g_{uu'} g_{vv'} \xi_x \xi_y \xi_z \,
d (\xi^z \omega^{uv}) \wedge \omega^{u'x} \wedge \omega^{v'y} \nonumber \\  
&- (2 \ell)^{-2} \phi \epsilon_{uvwxz} \xi^z \xi_{x'} \xi_{y'} g_{w' y} \, \omega^{uv} \wedge \omega^{ww'} \wedge \omega^{xx'} \wedge \omega^{yy'} \nonumber \\  
&- 2^{-1} h \phi^{-3} \epsilon_{uvxyz} \xi^z  \xi_{x'} \xi_{y'} \,  
d \xi^u \wedge d \xi^v \wedge \omega^{xx'} \wedge \omega^{yy'} + \lambda^+,  
\end{eqnarray}
where
\begin{eqnarray} \label{LagXi}
&\lambda^+ =  (\kappa' - \kappa) \phi^{-1} \beta^u \xi^v \xi_w d \xi^w \wedge \eta_{uv}(\xi)  + \kappa'  \phi \beta^v  d \xi^u\wedge \eta_{vu}(\xi) \nonumber \\
&+ 2^{-1} \kappa \phi^2 \beta^u \beta_u \eta(\xi)  - \hat \Lambda \phi^4 \eta(\xi).
\end{eqnarray}

The scalar field $\phi$ is defined as a function of the fields $\xi_u$ by eq.\ (\ref{Aug}) and we have introduced the covariant forms
\begin{equation} \label{EtaICov}
\eta_{vw}(\xi) = - 6^{-1}  \phi^{-3} \epsilon_{vwxyz} \xi_{x'} \xi_{y'} \xi_{z'} \,
\omega^{xx'} \wedge \omega^{yy'} \wedge \omega^{zz'},
\end{equation}
\begin{equation} \label{EtaCov}
\eta(\xi) = 2^{-3}  \omega^{vw} \wedge \eta_{vw}(\xi).
\end{equation}

This Lagrangian is pseudoscalar, because each term contains an odd number of pseudo-tensors $\omega^{uv}$, $\epsilon_{uvwxz}$ or $\beta_u$.  Besides the modified cosmological term proportional to $\hat \Lambda$, it contains only one constant with non trivial dimension, namely the action $\ell^2$.  It also contains the  adimensional parameters  $\kappa$ and $\kappa'$, that, as we shall see,  determine the coupling of the dilatonic and augmentonic  fields with matter.  We shall see in the next Section that, in order to obtain a reasonable theory, we have to choose a suitable value of $h$.

As in Section \ref{Revisited}, we assume a minimal coupling of matter with geometry, namely that the matter Lagrangian $\lambda^M$ does not contain structure coefficients,  the fields $\beta_u$ and the  derivatives of $\beta_u$, $\xi_u$  and $\phi$.  The conjugate forms are 
\begin{equation}  \label{ConjAug1}
\sigma_{uv} = - \ell^{-1} \phi g_{uw}  g_{vz} \xi_x \xi_y \,
\omega^{wx} \wedge \omega^{zy},
\end{equation}
\begin{eqnarray}  \label{ConjAug2}
& \pi^{\xi}_u = (- 2^{-1} \phi^{-2} \xi_u \, \omega^{xy} + h \ell \phi^{-4} \epsilon_u{}^{wxyz} \xi_z  \,  d \xi_w) \wedge \sigma_{xy}  \nonumber \\ 
&+ (\kappa - \kappa') \phi^{-1} \xi_u \beta^w \xi^v \eta_{wv}(\xi) +\kappa' \phi \beta^v \eta_{vu}(\xi),  
\end{eqnarray}

The  form conjugated to  $\beta_u$ vanishes and the corresponding field equation is
\begin{equation} \label{TanBeta2}
\frac{\partial \lambda}{\partial \beta^u} = (\kappa - \kappa') \xi^v d \phi  \wedge \eta_{uv}(\xi) + \kappa' \phi d \xi^v\wedge \eta_{uv}(\xi) 
+\kappa \phi^2 \beta_u \eta(\xi)  = 0.
\end{equation}
The field  $\xi_u$ satisfies the equation
\begin{equation}  \label{EqAug}
\frac{\partial \lambda}{\partial \xi^w} =  d \pi^{\xi}_w,
\end{equation} 
that is treated with more detail in the next Section.

The field equations can be simplified using a covariance transformation to choose  a suitable basis in the space $\mathcal{T}$. Since the covariance group is a global symmetry, in general one can simplify the equations only at a given point $s \in \mathcal{S}$.  We say that we have chosen an ``adapted basis'' at $s$.

We choose this basis in such a way that, at the given point $s$, the field $\xi_u$ satisfies the conditions (\ref{Special2}). We do not use the covariance under total dilatations to fix the value of $\phi$.  Note that by means of Lorentz transformations or total dilatations one can find other adapted bases at the same point.  

In an adapted basis, we have
\begin{equation} \label{EtaAd}
\eta_{i4}(\xi) = - \eta_{4i}(\xi) = \eta_i, \qquad \eta_{44}(\xi) = \eta_{ik}(\xi) = 0, \qquad \eta(\xi) = \eta
\end{equation}
and the field equation (\ref{TanBeta2}), for $\kappa, \kappa' \neq 0$,  takes the form
\begin{equation}  \label{TanBeta4}
A_{[mn]} \xi_u = A_{[mn]} \phi = 0,  \qquad 
A_i \phi  = \phi \beta_i ,  \qquad  \kappa'  A_i \xi^i = \kappa  \phi \beta_4.
\end{equation}
 
We also have, in the Lorentz tensor formalism,
\begin{eqnarray}  \label{ConjAug3}
&\sigma_i =\sigma_{i4} = 0, \qquad 
\sigma_{ik}  = - \ell^{-1}  \phi^3  g_{ij}  g_{kl} \, \omega^j \wedge \omega^l, \nonumber \\
&\sigma_{[ik]} = 2^{-1} \phi^3 \epsilon_{ikjl} \, \omega^j \wedge \omega^l,  
\end{eqnarray}
\begin{eqnarray}  
&\pi^{\xi}_i = 2 h \phi^{-3}d \xi^k \wedge \sigma_{[ik]} +\kappa' \phi \beta_4 \eta_i
= - 2 h (A_i \xi^k - \delta_i^k A_j \xi^j ) \eta_k + \kappa' \phi \beta_4 \eta_i,  \nonumber \\
&\pi^{\xi}_4 = - (2 \phi)^{-1} \, \omega^{[ik]} \wedge  \sigma_{[ik]}  + \kappa \phi \beta^i \eta_i.
\end{eqnarray}

The Lagrangian  $\lambda^G$ in an adapted basis becomes
\begin{eqnarray}  \label{LambdaBar}
&\lambda^G =  2^{-1} d \omega^{[ik]} \sigma_{[ik]}  + \overline \lambda^G,  \nonumber \\
&\overline \lambda^G  =  2^{-2} \phi^2 \epsilon_{ikjl} 
(\phi^2 g_{mn} \, \omega^{[im]} \wedge \omega^{[nk]} 
+ d \phi \wedge \omega^{[ik]}) \wedge  \omega^j \wedge \omega^l  \nonumber \\
&+ h ( A_i \xi^i A_k \xi^k - A_i \xi^k A_k \xi^i) \eta + \lambda^+,   
\end{eqnarray}
\begin{equation} \label{LagXi2}
\lambda^+ = - 2^{-1} \kappa \phi^2 \beta^u \beta_u \eta - \hat \Lambda \phi^4 \eta. 
\end{equation}
We have used the equations  (\ref{TanBeta4}).

In order to write the field equation (\ref{Tangential2}) in an adapted basis, we have to compute first the exterior derivative $d \sigma_{\alpha}$ and then to impose the conditions (\ref{Special2}). In this way we obtain
\begin{eqnarray} \label{DSigma}
&d \sigma_i = \ell^{-1}  \phi^2 \epsilon_i{}^{jlk} A_j \xi_l \, \eta_k, \nonumber \\
&d \sigma_{[ik]} = \phi^3 \epsilon_{ikjl} \,  d \omega^j \wedge \omega^l 
- 3 \phi^3 (\beta_i \eta_k  - \beta_k \eta_i) \nonumber \\
&- \ell \phi^2 A_q \xi_p (\delta_i^p g_{km} g_{ln} + \delta_l^p g_{im} g_{kn} + \delta_k^p g_{lm} g_{in}) \,  \omega^{[mn]} \wedge \omega^q \wedge \omega^l.
\end{eqnarray}
 
From eqs.\   (\ref{Tangential2}) and (\ref{TauAlphaS}), we have
\begin{eqnarray} \label{TauPar}
&- d \sigma_i -  2^{-1}  d \omega^{[jl]} \wedge i_{i} \sigma_{[jl]} + A_{i} \xi_u \pi_{\xi}^u - i_{i} \overline\lambda^G  = \tau_{i}^M, \nonumber \\
&- d \sigma_{[ik]}  -  i_{[ik]} \overline\lambda^G = \tau_{[ik]}^M, 
\end{eqnarray}
where $\tau_{\alpha}^M$,  given by eq.\ (\ref{TauM}), describes the 10-momentum of the matter fields, excluding $\xi_u$.

If we use the expressions (\ref{DSigma}), we obtain the field equations
\begin{eqnarray} \label{AugTan2}
&- 2^{-1} \phi^3 \epsilon_{ijkl} (d \omega^{[jk]} + \phi g_{mn} \, \omega^{[jm]} \wedge \omega^{[nk]}) \wedge \omega^l  \nonumber \\
&- 2^{-1}  \phi^3 \epsilon_{ijkl} \, \beta_p \, \omega^p \wedge \omega^{[jk]} \wedge \omega^l   =  T^{\xi k}_i \eta_k + \tau^M_i,
\end{eqnarray}
\begin{eqnarray} \label{AugTan1}
&-  \phi^3 \epsilon_{ikjl} (d \omega^j + \phi g_{mn} \, \omega^{[jm]} \wedge \omega^n) \wedge \omega^l  \nonumber \\
&+\ell \phi^2 A_q \xi_p (\delta_i^p g_{km} g_{ln} + \delta_l^p g_{im} g_{kn} + \delta_k^p g_{lm} g_{in}) \,  \omega^{[mn]} \wedge \omega^q \wedge \omega^l =   \nonumber \\
&T^{\xi j}_{[ik]} \eta_j + \tau^M_{[ik]}, 
\end{eqnarray} 
where the quantities
\begin{eqnarray} \label{TXi}
&T^{\xi k}_i =  2 h (A_i \xi^j A_j \xi^k - A_i \xi^k A_j \xi^j) 
- h (A_j \xi^l A_l \xi^j - A_j \xi^j A_l\xi^l) \delta_i ^k   \nonumber \\
&+ \ell^{-1}  \phi^2 \epsilon_i{}^{jlk} A_j \xi_l +  T^{+ k}_i,
\end{eqnarray}
\begin{equation} \label{TPlus}
T^{+ k}_i = \kappa \phi^2 (\beta_i \beta^k - 2^{-1}\beta_u \beta^u\delta_i^k)
- \kappa' \phi \beta_4  A_i \xi^k
- \hat \Lambda  \phi^4 \delta_i ^k, 
\end{equation}
\begin{equation} \label{TXi2}
T^{\xi j}_{[ik]} =  - 2 \phi^3  (\beta_i \delta^j_k  -  \beta_k\delta^j_i )
\end{equation}
may be interpreted as the 10-momentum of the augmentonic and  dilatonic fields. If $\xi_i = 0$ everywhere, these equations coincide with the  geometric equations (\ref{GBD1}) and  (\ref{GBD2}) of the GBD theory.  

By considering separately the terms containing different products of the forms $\omega^i$ and $\omega^{[ik]}$ and assuming the spatially localized form (\ref{Localized}) for the 10-momentum of matter, we obtain 
\begin{eqnarray}  \label{HorDiff5}
&F_{[jl][mn]}^{[ik]} = \phi \hat F_{[jl][mn]}^{[ik]}, \qquad   F_{p[jl]}^{[ik]} 
= \beta_p (\delta_j^i \delta_l^k - \delta_l^i \delta_j^k),  \nonumber \\
&\phi^3 (- F_{jk}^{[ik]} + 2^{-1} \delta^i_j F_{lk}^{[lk]}) = T^{\xi i}_j  + T^i_j, 
\end{eqnarray}
\begin{eqnarray} \label{VerDiff5}
 &F_{[jk][mn]}^i = 0, \qquad  F_{[jk]l}^i = \phi \hat F_{[jk]l}^i 
-  \ell \phi^{-1} \epsilon_{jk}{}^{pi} A_l \xi_p, \nonumber \\
&\phi^3 (F_{jk}^i + \delta^i_j F_{kl}^l - \delta^i_k F_{jl}^l) 
- 2 \phi^3 (\beta_k \delta^i_j - \beta_j \delta^i_k) = T^i_{[jk]}.
\end{eqnarray} 

From these equations we have the useful formulas
\begin{eqnarray} \label{Useful3}
&d(\phi^{-1} \xi_u \omega^{4u}) = - \phi^{-1} \,  d \xi_k  \wedge \omega^k, \nonumber \\
&d(\phi^{-1} \xi_u \omega^{iu}) = - 2^{-1} F^i_{jk} \, \omega^j \wedge \omega^k
- \phi g_{jk} \,  \omega^{[ij]} \wedge \omega^k,
\end{eqnarray}
\begin{eqnarray} \label{Useful4}
&d \eta_{i4}(\xi) =  -  F_{ik}^k \, \eta +  \phi g_{ij} \, \omega^{[kj]}  \wedge \eta_k, \nonumber \\
&d \eta_{ik}(\xi) =  \phi^{-1}  (A_i \xi_k - A_k\xi_i) \eta.
\end{eqnarray}

The equations (\ref{PhiDef}) and (\ref{Beta2}) of the GBD theory, that give $\phi$ and $\beta_i$ as functions of the structure coefficients, are still valid in an adapted basis and from eq.\ (\ref{Special2}) we obtain all the components of $\xi_u$.  From eqs.\  (\ref{HorDiff5}) and (\ref{VerDiff5}) we also find
\begin{eqnarray}  \label{BetaGeom}
&F_{i[jk]}^{[jk]} = 12 \beta_i,  \qquad 
F_{[ik] l}^l = \ell \phi^{-1} \epsilon_{ik}{}^{jl} A_j \xi_l ,  \nonumber \\
&\epsilon^{jk}{}_{im}  F_{[jk]n}^i = - 6 \ell \phi^{-1} A_n \xi_m, \qquad
\epsilon^{jkn}{}_i  F_{[jk]n}^i =  6 \ell \kappa (\kappa')^{-1} \beta_4.
\end{eqnarray}

If the spinning matter is described by Dirac fields, we have
\begin{equation} \label{DiracSpin}
T_{[jk]}^i = \epsilon_{jk}{}^{il}W_l
\end{equation}
and from eq.\ (\ref{VerDiff5}) we have
\begin{equation} \label{TorDir}
F_{ji}^i  = 3 \beta_j,  \qquad
F_{jk}^i =  \phi^{-3} \epsilon_{jk}{}^{il}W_l - \beta_k  \delta^i_j +  \beta_j  \delta^i_k. 
\end{equation}

\section{The conservation laws}
\label{Conservation} 

We assume that all the field equations, apart from the equation (\ref{EqAug})
for the augmentonic field, are satisfied and that
\begin{equation}  \label{Cond}
d \theta_{td} = 0,  \qquad   \xi^v d \theta_{uv} = 0.
\end{equation} 
If we remember that the action of the infinitesimal transformations $X_{td}$ and $X_{uv}$ is given by 
eqs.\  (\ref{TotalDil}),  (\ref{XVWF}), (\ref{XI4Omega}) and
\begin{equation} 
X_{td} \xi_u  = \xi_u,
\end{equation}  
from eq. (\ref{Invariance2}) we have
\begin{equation} 
\xi_w \left( \frac{\partial \lambda}{\partial \xi_w} - d \pi_{\xi}^w \right) = 0, \qquad
\xi^v (g_{vw} \xi_u - g_{uw} \xi_v) \left( \frac{\partial \lambda}{\partial \xi_w} - d \pi_{\xi}^w \right) = 0.
\end{equation}   
and the missing eq.\  (\ref{EqAug})  follows. 

We have shown that, if all the other field equations are satisfied, eq.\   (\ref{EqAug}) is equivalent to eq.\ (\ref{Cond}), in particular it follows from the conservation laws for $\theta_{td}$ and $\theta_{uv}$. In an adapted basis, the second condition in eq.\  (\ref{Cond}) takes the simpler form  $d \theta_{i4} = 0$.

The geometric contributions to the conserved forms defined by eq. (\ref{Conserved}) are given by
\begin{eqnarray} \label{ThetaTDAug}
&\theta^G_{td} = - 2^{-1} \, \omega^{uv} \wedge \sigma_{uv} + \xi_u \pi_{\xi}^u = \nonumber \\
&\kappa \phi \beta^u \xi^v \eta_{uv}(\xi),
\end{eqnarray}
\begin{eqnarray} \label{ThetaUVAug}
&\theta^G_{uv} = (\delta_u^x g_{vz} - \delta_v^x g_{uz}) \, \omega^{zy} \wedge \sigma_{xy}
+ \xi_v \pi^{\xi}_u - \xi_u \pi^{\xi}_v \nonumber \\
&= ((\delta_u^x g_{vz} - \delta_v^x g_{uz}) \, \omega^{zy} 
+ 2^{-1} h \ell \phi^{-3} \epsilon_{uv}{}^{xyz} (\phi d \xi_z - \xi_z  d \phi)) 
\wedge \sigma_{xy} \nonumber \\
&+ \kappa' \phi \beta^w (\xi_v \eta_{wu}(\xi) - \xi_u \eta_{wv}(\xi)).
\end{eqnarray}

In an adapted basis we have
\begin{equation} \label{ThetaTDAug1}
\theta^G_{td} = - \kappa \phi^2 \beta^i \eta_i,
\end{equation}
\begin{equation} \label{ThetaUVAug1}
\theta^G_{i4} =   - h \phi (A_i \xi^k - A_j \xi^j \delta_i^k ) \eta_k + \kappa' \phi^2 \beta_4 \eta_i.
\end{equation}
If $\xi_i = 0$ everywhere, the second expression vanishes and the first one coincides with eq.\ (\ref{ThetaDil}) of the GBD theory.  In conclusion, we have seen that, in the absence of matter, the solutions of the GBD theory are also particular solutions of the theory we are considering.  It follows that we can use eq.\ (\ref{KOmega}) to find the value of the constant $\kappa$.

Note that, unless the condition (\ref{Special2}) is valid everywhere, the last equations cannot be used to write the conservation laws in differential form, since, as we have already remarked, we have to perform the exterior differentiation before applying this condition.  In this way, we obtain
\begin{eqnarray} \label{DThetaTdAug}
&d \theta_{td}^G = \kappa \phi^2 (- A_i \beta^i + \beta_i \beta^i ) \eta \nonumber \\
&- 2^{-1} \kappa\ell \phi \beta_i A_k \xi_j \epsilon^{ij}{}_{mn}  \ \omega^{[mn]}  \wedge \eta^k.
\end{eqnarray}

This equation is the same as the eq.\ (\ref{GBD4}) of the GBD theory, with a new terms containing  $\omega^{[mn]}$.  If $d \theta_{td}^M$ does not contain $\omega^{[mn]}$, this term has to vanish and we obtain the equation
\begin{equation}  \label{Condition1}
\beta_i A_k \xi_j - \beta_j A_k \xi_i = 0.
\end{equation}
 
From the form (\ref{ThetaUVAug}), after a long calculation also using eqs.\ (\ref{Useful3}) and (\ref{Useful4}), we find
\begin{eqnarray} \label{DThetaI4Aug}
&d \theta_{i4}^G =  - h \phi (\beta_i A_p \xi^p - \beta_p A_i \xi^p) \eta
+ \phi (A_i \xi^l F^m_{ml} + A_m \xi^l F^m_{il} 
+ A_l \xi^l F^m_{im}) \eta \nonumber \\
&- \kappa' \phi  (\beta^k A_i \xi_k - \phi A_i \beta_4 - 2 \phi \beta_4 \beta_i 
+ \phi \beta_4 F_{ik}^k) \eta  \nonumber \\
&+ (1 - h) \phi^2 A_k \xi^p (\delta_p^k g_{mi} \delta_n^q 
+ \delta_m^k g_{ni} \delta_p^q - \delta_i^k g_{mp} \delta_n^q 
- \delta_m^k g_{np} \delta_i^q) \, \omega^{[mn]}  \wedge \eta_q \nonumber \\
&+ 2^{-1}\kappa' \phi^2  (A_{[mn]} \beta_4 \delta_i^q 
- \phi \beta_4 (g_{in} \delta_m^q  -  g_{im} \delta_n^q)) \, \omega^{[mn]}  \wedge \eta_q.
\end{eqnarray} 

 If the matter contribution $d \theta_{i4}^G$ does not contain the forms $\omega^{[mn]}$, also the terms of eq.\ (\ref{DThetaI4Aug}) containig these forms must vanish and we obtain the equation
\begin{eqnarray} \label{DThetaI4Aug2}
&(1 - h)  (A_k \xi^k  g_{mi} g_{nq} 
+ A_m \xi_q  g_{ni} - A_i \xi_m  g_{nq} - A_m \xi_n g_{iq} \nonumber \\
&- A_k \xi^k  g_{ni} g_{mq} 
- A_n \xi_q  g_{mi} + A_i \xi_n  g_{mq} + A_n \xi_m g_{iq}) \nonumber \\
&+\kappa'   (A_{[mn]} \beta_4 g_{iq} 
- \phi \beta_4 (g_{in} g_{mq} -  g_{im} g_{nq})) = 0.
\end{eqnarray} 

If we contract the indices $i$ and $q$ we have
\begin{equation} \label{DThetaI4Aug3}
2 \kappa'  A_{[mn]} \beta_4 = (1 - h) (A_m \xi_n  -  A_n \xi_m) 
\end{equation} 
and, if we substitute this expression into eq.\ (\ref{DThetaI4Aug2}), we have
\begin{eqnarray} \label{DThetaI4Aug4}
&(1 - h)  (A_k \xi^k  g_{mi} g_{nq} 
+ A_m \xi_q  g_{ni} - A_i \xi_m  g_{nq} - 2^{-1} A_m \xi_n g_{iq} \nonumber \\
&- A_k \xi^k  g_{ni} g_{mq} 
- A_n \xi_q  g_{mi} + A_i \xi_n  g_{mq}  + 2^{-1} A_n \xi_m g_{iq}) \nonumber \\
&+ \kappa'  \phi \beta_4 (g_{in} g_{mq} -  g_{im} g_{nq}) = 0.
\end{eqnarray} 

If we contract the indices $q$ and $n$ and we use eq.\ (\ref{TanBeta4}), we obtain the equation
\begin{equation} \label{DThetaI4Aug5}
(2 (1 - h) - 3 \kappa^{-1} (\kappa')^2) A_k \xi^k  g_{im} 
+ 2^{-1} (1 - h) (A_m \xi_i - 5 A_i \xi_m) = 0.
\end{equation} 
If $h \neq 1$, 
it follows that
\begin{equation} 
A_m \xi_i  = 2^{-2} A_k \xi^k g_{im},
\end{equation} 
an equation that is too restrictive.  We also find
\begin{equation} 
h = 1- 2 \kappa^{-1} (\kappa')^2.
\end{equation} 

It is more interesting to satisfy eq.\ (\ref{DThetaI4Aug5}) assuming that 
\begin{equation} \label{Condition2}
 h =  1, \qquad A_k \xi^k = 0, \qquad  \beta_4 = 0.
\end{equation}  
It follows that 
\begin{equation} \label{Follows}
A_{[mn]} \beta_4 = 0, \qquad  \xi^u \beta_u = 0.
\end{equation} 
The last equation is covariant and is valid in a general basis.  Considering its derivative, we find, in an adapted basis,
\begin{equation} \label{Follows1}
\phi A_i \beta_4  = \beta_k A_i \xi^k.
\end{equation}

The equations (\ref{Condition2}), and  (\ref{Follows})  assure that the whole equation (\ref{DThetaI4Aug2}) is satisfied and that the terms in eq.\  (\ref{DThetaI4Aug})  that contain $\omega^{[mn]}$ cancel. Also the other terms proportional to $\kappa'$ cancel, as a consequence of eq.\  (\ref{Follows1}).  If we use eq.\ (\ref{TorDir} ) to calculate the terms containing torsion, we finaly obtain
\begin{equation} \label{DThetaI4Aug6}
d \theta_{i4}^G = - 3 \phi \beta_k A_i \xi^k \eta + \phi^{-2} W_p  \epsilon_{il}{}^{mp} A_m \xi^l = - d \theta_{i4}^M. 
\end{equation} 

Since the covariance transformations act on the spin indices of the matter fields, the forms $\theta_{i4}^M$, as well as the quantities $W_p$, depend on  Fermionic fields and require a quantum treatment, that will be discussed  elsewere.  In the absence of spinning matter, we have
\begin{equation} \label{DThetaI4Aug7}
d \theta_{i4}^G = - 3 \phi \beta_k A_i \xi^k \eta = - 3 \phi^2  A_i \beta_4 \eta = 0.
\end{equation}

\section{A foliation of $\mathcal{S}$}
\label{Foliation}
The theory we have presented is formulated in the space $\mathcal{S}$ with a structure of absolute parallelism, but, as a consequence of the field equations,  it maintains some features (not all) of the structure of principal bundle present, for instance, in the EC theory.

If, at a given point $s \in \mathcal{S}$, we choose an adapted basis in $\mathcal{T}$, we can define in the corresponding tangent space $\mathcal{T}_s$ a generalized vertical subspace $\mathcal{T}_{Vs}$ generated by the vectors $A_{[ik]}$.  The adapted basis is defined up to a Lorentz transformation, but this ambiguity does not affect the definition of $\mathcal{T}_{Vs}$.  Remember that there is no covariantly defined vertical subspace of $\mathcal{T}$.

If we apply this procedure to all the points of $\mathcal{S}$, we obtain a distribution of subspaces of the tangent spaces described by the differential system, covariant under $SO(2,3)^c_A$,
\begin{equation} \label{DiffSys}
\xi_v \omega^{uv} = 0.
\end{equation}

This distribution is integrable \cite{KN,CB} if
\begin{equation} \label{IntDis}
d (\xi_v \, \omega^{uv}) = 0
\end{equation}
is a consequence of eq.\ (\ref{DiffSys}).  One can easily compute this exterior derivative in an adapted basis,  in which eq.\ (\ref{DiffSys}) takes the simple form $\omega^i = 0$.  By means of  the field equations we see that every term contains a factor of the kind $\omega^i$ and the required condition is satisfied.

We can apply  Frobenius' theorem \cite{KN,CB}, that assures that  every point is contained in 6-dimensional open submanifold, called a ``leaf'', that is tangent, at all its points,  to the corresponding generalized vertical subspace. In this way, we define a generalized kind of local space-time coincidence.  It follows from eq.\ (\ref{TanBeta4}) that the quantities $\xi_u $ are constant on each leaf.

One can consider the leaves as the points of a space-time that, however, is not necessarily a manifold. It may be useful to introduce, tentatively, further assumptions of a global nature, namely that the leaves are the fibers of a fiber bundle over a 4-dimensional space-time manifold.  However,  a full structure of principal fibre bundle would contradict some of the field equations, if the fields $\xi_u$ are not constant.  Also the introduction of a pseudo-Riemannian metric tensor on the base manifold is, in general, not possible

\section{Open problems and work in progress}
\label{Conclusions} 

The theory with partially augmented covariance treated in the preceding Sections requires several further investigations, both from the theoretical and the phenomenological points of view.  

For instance, one has to consider its applications to cosmology.  In particular, one should see if the augmentonic fields, including the BD field, may help to replace the dark energy and the dark matter that have been introduced to explain the accelerated  cosmic expansion in the last few billion years  \cite{Riess,Perlmutter}. The description of earlier ages introduces more difficult problems, that will be treated in a future invastigation within the framework of fully covariant theories.

Cosmological models including a scalar field have been analyzed by many authors  (see for instance \cite{Faraoni}).  An analysis including also a 4-vector field can be found in ref.\ \cite{Moffat,Moffat2}.  

It is interesting to remark that a relevant cosmological effect of the BD field $\phi$ is possible in spite of the very small value of its derivative \cite{MU}, because the value of the constant $\kappa$ (or of the BD parameter $\omega$) is very large \cite{Will}.  It is not clear that a similar compensation can take place when we consider the contribution of the augmentonic field $\xi_0$ (see eq.\ (\ref{TPlus})).

A careful discussion of the propagation of light in the modified geometric background could be relevant for the interpretation of the astronomical observations.

One should also consider a possible application to galactic dynamics, in particular to the explanation of the rotation curves of the peripheral stars \cite{ST}.

In dealing with these applications, it may be necessary to consider possible generalizations of the Lagrangian (\ref{LagAug}), that is not univocally determined by its covariance property. 

In any case, one has to develop more powerful techiques to treat solutions symmetric with respect to a given group of diffeomorphisms, for instance homogeneous  isotropic solutions or static spherically symmetric solutions.  Some progress in this direction will appear soon.

Both in view of the applications and of the formulation of new theoretical ideas, the next step is to look for a Lagrangian invariant under the fully augmented covariance group $GL(4, \mathbf{R})$, solving the problems indicated in Section \ref{Partially}.  Remember that the Lagrangian (\ref{LagAug}) is a generalization of the GBD Lagrangian (\ref{GBDLag}), that, in turn, generalizes the NR Lagrangian (\ref{NR}).

Perhaps, the most interesting aspect of this development could be the appearance of a conserved form $\theta_{45} =\theta(X_{45})$ (see eq.\ (\ref{Conserved})), where $X_{45}$ generates rotations involving the components $f_4$  and $f_5$ of a 6-vector.  

The corresponding integrated quantity $Q_{45}$ (see eq.\ (\ref{ConservedQ})) is invariant under the  Lorentz group and under total dilatations. In all the Lorentz invariant vacuum states  it takes the same value, that, for dimensional reasons,  must have the form $\alpha' \ell^2$ with a numerical coefficient $\alpha' $ given by the theory.

In analogy with Bohr's atomic theory, we  have to require $Q_{45}  =  n \hbar$, where $n$ is an integer and,  if $\alpha' \neq 0$, we find the possible discrete values of $\ell^2 / \hbar$ (see the left arrow of the triangle (\ref{Triangle})). 

Another interestig development is the introduction of a supercovariance group.  We have suggested in Section \ref{Partially} that a possible candidate is the super Lie group $OSp(N | 4; \mathbf{R})$, that contains $Sp(4, \mathbf{R})$ and the compact internal covariance group $SO(N, \mathbf{R})$.  

From this investigation we might obtain a relation between $\ell$, that appears in the definition of $Sp(4, \mathbf{R})$,  and the coupling strength $g$ of a grand unification theory of particle interactions involving a subgroup of  $SO(N, \mathbf{R})$   (see the right arrow of the triangle (\ref{Triangle})).

If this program has success, since the ratio $g^2/\hbar$ represented by the lower arrow of  the triangle (\ref{Triangle}) is approximately known, we could obtain some information about the value of $\ell$ and of the quantum number $n$.

\section*{Acknowledgments}

It is a pleasure to thank G. Cognola, L. Vanzo and S. Zerbini, who have participated, several years ago, to the beginning of this research program and, also more recently, have helped with fruitful advices and discussions. I also like to thank prof. F. Hehl for a stimulating correspondence.

\end{document}